\documentclass[epj]{svjour}
\usepackage{graphics}
\usepackage{booktabs}

\usepackage{amsmath}
\usepackage{amssymb,amsmath}
\usepackage{mathrsfs}
\usepackage{color}
\usepackage[margin=1 in]{geometry}

\pdfoutput=1

\usepackage{enumitem}
\usepackage{enumerate}
\usepackage{hyperref}

\numberwithin{equation}{section}

\renewcommand{\phi}{\varphi}

\usepackage{bm}

\begin{document}
\title{New classes of quadratically integrable systems with velocity dependent  potentials: non-subgroup type cases}
\author{Md Fazlul Hoque https://orcid.org/0000-0001-8427-1489 \inst{1,3}\thanks{\emph{email:} fazlulmath@pust.ac.bd} \and Ond\v{r}ej Kub\r{u} https://orcid.org/0000-0001-7463-5680 \inst{1}\thanks{\emph{email:} ondrej.kubu@fjfi.cvut.cz} \and Antonella Marchesiello https://orcid.org/0000-0002-1134-7478 \inst{2}\thanks{\emph{email:} marchant@fit.cvut.cz} \and Libor \v{S}nobl https://orcid.org/0000-0002-7270-6251 \inst{1}\thanks{\emph{email:} libor.snobl@fjfi.cvut.cz (corresponding author)}
%
} 
%
%
\institute{Czech Technical University in Prague, Faculty of Nuclear Sciences and Physical
	Engineering, Department of Physics, B\v{r}ehov\'{a} 7, 115 19 Prague 1, Czech Republic \and Czech Technical University in Prague, Faculty of Information Technology,
	Department of Applied Mathematics, Th\'{a}kurova 9, 160 00 Prague 6, Czech Republic
\and Pabna University of Science and Technology, Faculty of Science, Department of Mathematics, Pabna 6600, Bangladesh}
\date{Received: date / Revised version: date}
%
\abstract{	
We study quadratic integrability of systems with velocity dependent potentials in three-dimensional Euclidean space. Unlike in the case with only scalar potential, quadratic integrability with velocity dependent potentials does not imply separability in the configuration space. The leading order terms in the pairs of commuting integrals can either generalize or have no relation to the forms leading to separation in the absence of a vector potential. We call such pairs of integrals generalized, to distinguish them from the standard ones, which would correspond to separation. Here we focus on three cases of generalized non-subgroup type integrals, namely elliptic cylindrical, prolate / oblate spheroidal and circular parabolic integrals, together with one case not related to any coordinate system. We find two new integrable systems, non-separable in the configuration space, both with generalized elliptic cylindrical integrals. In the other cases, all systems found were already known and possess standard pairs of integrals. In the limit of vanishing vector potential, both systems reduce to free motion and therefore separate in every orthogonal coordinate system.
\PACS{
 {02.30.Ik}{Integrable systems} \and
{03.50.De}{Classical electromagnetism, Maxwell equations}
} 
} 
\maketitle

\section{Introduction}
This paper is a contribution to the classification of integrable systems with velocity dependent potentials on the three-dimensional (3D) Euclidean space $\mathbb{E}_3$.
Let us recall that a 3D classical Hamiltonian system is integrable if it admits three functionally independent integrals (constants of motion) that are in involution with respect to the Poisson bracket, including the Hamiltonian. If there exist more than three independent integrals, out of which three are in involution, the system is said to be superintegrable. 

For natural systems, i.e. systems whose Hamiltonian reads $H=T_2+T_0$, where $T_2$ is the kinetic term quadratic in the momenta and $T_0$ is the scalar potential, the classification of integrable system with integrals that are at most second order polynomials in the momenta is well known and in \cite{Makarov1967} it was shown that each pair of commuting quadratic integrals is related to an orthogonal system of coordinates in which the Hamilton-Jacobi equation separates (for the simpler 2D case  see~\cite{FrisMandrosov,WinternitzSmorodinski}). Superintegrability then corresponds to multi-separability, that is separability in more than one coordinate system \cite{FrisMandrosov,Makarov1967,Evans}. 

The situation is more involved when the Hamiltonian contains also terms linear in the momenta, due e.g. to the presence of a magnetic field or a particle considered in a rotating frame. In this case, separation of the Hamilton-Jacobi equation in the configuration space (or, in quantum context, of the Schr\"odinger equation) is possible only if one first order integral exists \cite{BaShaMe,Benenti_2001}. Thus, the relation between quadratic integrability and separability is lost and also the structure allowed for the leading order terms of the integrals becomes more general \cite{Marchesiello2022}. Investigation of systems with velocity dependent potentials started in two dimensions in~\cite{DoGraRaWin} and was further pursued in~\cite{McSWin,BeWin,Pucacco_2005}. A systematic search for quadratically integrable systems with the most general allowed structure of integrals  has started in \cite{KMS2022}, where all integrable systems admitting generalized cylindrical type integrals have been found and furthermore it was proven that integrable systems with generalized spherical type integrals cannot exist. Such integrals are so called as their leading order terms generalize the structure which would correspond to separation in cylindrical and spherical coordinates without magnetic field, respectively, by inclusion of additional terms. Furthermore, next to the cases where the integrals can be related to the ones corresponding to a coordinate system in which the Hamilton-Jacobi equation would separate (in absence of magnetic field), new classes of integrable systems can exist \cite{Marchesiello2022} with second order terms of the integrals apparently not related to any of the ones listed in \cite{Makarov1967}. For concrete examples of such systems see~\cite{Marchesiello2022,HS2022}.

In this work we continue our study of integrable systems with generalized structure of the integrals, aiming at a complete classification of quadratically integrable systems. Some disparity has been observed between subgroup and non-subgroup type integrals, already in the absence of magnetic field \cite{Marchesiello2015a}. In particular, Cartesian and cylindrical type integrals, related to the two maximal Lie subalgebras of the 3D Euclidean algebra, seem to play a prominent role in the classification theory of integrable systems \cite{Miller2013,KMS2022}. To shed light on this point on one hand and to advance in the classification problem on the other hand, in this work we consider the conditions for the existence of three non-subgroup type generalized pairs of integrals, namely elliptic cylindrical, prolate / oblate spheroidal and circular parabolic integrals, together with the possible existence of systems with a pair of integrals with quadratic terms not related to any class corresponding to separation (in absence of magnetic field). We found two new classes of integrable systems in the elliptic cylindrical case and none in any of the other cases. 

For all other integrable systems known so far with generalized structure of the integrals, the limit for vanishing magnetic field leads to separation in cylindrical~\cite{KMS2022} or Cartesian coordinates~\cite{Marchesiello2017}, even when the leading order terms of the integrals do not correspond to such coordinate systems \cite{Marchesiello2022,HS2022}. In the limit for vanishing magnetic field, both new systems presented here reduce to free motion, thus trivially separate in every  orthogonal coordinate system, not just in cylindrical nor elliptic cylindrical coordinates, as one would expect. Separation in non--subgroup type coordinates seems not to arise in the limit of vanishing magnetic field when generalized non--subgroup type integrals are considered, unless the limit leads to free motion. 

The paper is organized as follows. In the following section \ref{sec:formulation} we introduce the system and the structure of the integrals we are interested in. In section \ref{sec:cylell} we present the determining equations for the existence of generalized elliptic cylindrical type integrals and introduce the two new integrable systems found. The details of the computation are given in Appendix \ref{sec:appendix}, for reader's convenience. The results for the other classes of integrals are briefly described in section \ref{sec:others}, as the computation can be done essentially by the same methods used in section \ref{sec:cylell} and in Appendix \ref{sec:appendix}. Some conclusions and perspectives are discussed in section \ref{sec:concl}.

\section{The system and the integrals}\label{sec:formulation}
Let us consider a classical charged particle in an external electromagnetic field in three dimensional Euclidean space $\mathbb{E}_3$. 
We choose the units of measurement so that $m=1$, $e=-1$. 

Let $(\bm{q}, \bm{p})\in\mathbb{R}^6$ be the canonical phase space coordinates and let $ H(\bm{q}, \bm{p})$ be the Hamiltonian of the system. We find it convenient to express the vector potential $A(\bm{q})$ and magnetic field $\textbf B(\textbf q)$ in the form of differential 1- and 2-forms, respectively. Namely,
\begin{equation}\label{A1form}
	A(\bm{q})=A_1(\bm{q})\mathrm{d}q_1+ A_2(\bm{q})\mathrm{d}q_2+A_3(\bm{q})\mathrm{d}q_3,
\end{equation}
and we obtain the magnetic field 2-form as the exterior derivative of the vector potential, i.e. $B=\mathrm{d}A$. It reads
\begin{eqnarray}\label{mag2form}
	B& =&\left( \frac{\partial A_3}{\partial q_2}- \frac{\partial A_2}{\partial q_3}\right) \mathrm{d}q_2 \wedge \mathrm{d}q_3 +\left( \frac{\partial A_1}{\partial q_3}- \frac{\partial A_3}{\partial q_1}\right)\mathrm{d}q_3 \wedge \mathrm{d}q_1 + \left( \frac{\partial A_2}{\partial q_1}- \frac{\partial A_1}{\partial q_2}\right) \mathrm{d}q_1 \wedge \mathrm{d}q_2 \nonumber\\
	&=& B_1(\textbf q) \mathrm{d}q_2 \wedge \mathrm{d}q_3+ B_2(\textbf q) \mathrm{d}q_3 \wedge \mathrm{d}q_1+ B_3(\textbf q) \mathrm{d}q_1 \wedge \mathrm{d}q_2.
\end{eqnarray}
Let $p_i^A=p_i+A_i(\textbf q)$ be the momenta covariant with respect to a time-independent gauge transformation
\begin{equation}\label{gauge}
	A'(\bm{q})= A (\bm{q})+\mathrm{d} \chi(\bm{q}), \quad V'(\bm{q})=V(\bm{q}),
\end{equation} 
where $\chi(\bm{q})$ is a smooth scalar function. In gauge covariant form the Hamiltonian of the system with magnetic field reads
\begin{equation}\label{Hamiltonian}
	H(\bm{q},\bm{p})=\frac12\sum_{j=1}^3 (p_j^A)^2+ W(\bm{q}).
\end{equation}
Notice that for a particle in a rotating frame the structure of the Hamiltonian would be formally the same as in~\eqref{Hamiltonian} and the field~\eqref{mag2form} would correspond to the angular velocity. This correspondence is successfully employed e.g in  the study of (super)integrability and separability of systems with constant magnetic fields~\cite{BKS2020,Zhang}.

We consider here the problem of the integrability of the Hamiltonian \eqref{Hamiltonian} under the assumption that a pair of commuting integrals $X_1, X_2$ exists, both polynomial in the momenta of at most second order. 
In the three dimensional Euclidean space, the highest order terms of any second order integral are polynomials of order $2$ in the enveloping algebra of the Euclidean (Poisson--) Lie algebra $\mathfrak{e}(3)$ with the basis ${p_1, p_2, p_3, \ell_1, \ell_2, \ell_3}$, where $\ell_j=\sum_{k,l} \epsilon_{jkl} q_k p_l$ are the angular momenta, $j=1,2,3$. Namely, in terms of covariantized momenta $p_j^A$, the integrals $X_k$ can be expressed as \cite{Miller2013}
\begin{equation}\label{classintUEACart}
	\begin{split}
		X_k ={}&\sum_{i,j:\; i\leq j}\alpha_{ij}\ell_i^A \ell_j^A+ \sum_{i,j}\beta_{ij}p_i^A \ell_j^A+\sum_{i,j:\; i\leq j} \gamma_{ij}p_i^A p_j^A +\\
		&+\sum_j s_k^{q_j}(\bm{q}) p_j^A +m_k(\bm{q}),\;\;k=1,2,
	\end{split}
\end{equation}
and have to satisfy
\begin{equation}\label{commutH}
	\{ H, X_k\}=0, \;\; k=1,2 
\end{equation}
together with
\begin{equation}\label{commutX1X2}
	\{ X_1, X_2 \}=0,
\end{equation}
where $\{\cdot,\cdot\}$ denotes the Poisson bracket.

The functions $s_k^{q_j}$ and $m_k$ and the coefficients $\alpha_{ij}, \beta_{ij}, \gamma_{ij} \in\mathbb{R}$ have to be found, together with the magnetic field \eqref{mag2form} and the electrostatic potential $W$ in \eqref{Hamiltonian}, in order to solve the determining equations for the integrals coming from \eqref{commutH}-\eqref{commutX1X2}. Besides the unknown functions, such equations contain polynomial terms in the momenta of at most second order. By collecting the coefficients of each power in $p_i p_j$ and equating them to zero, we see that each Poisson bracket in \eqref{commutH}-\eqref{commutX1X2} corresponds to a system of $10$ determining equations. The explicit form of the equations coming from \eqref{commutH} is known \cite{Marchesiello2015}. The commutativity conditions \eqref{commutX1X2} for the integrals are quite cumbersome, we will write them explicitly order by order only when needed and after they are simplified by the solution of \eqref{commutH} for the special classes of integrals we are going to consider in the following. 

It is known~\cite{Marchesiello2022} that, up to Euclidean transformations, there exist $11$ classes of pairs of commuting quadratic elements in the universal enveloping algebra of Euclidean algebra, each in principle corresponding to a possible structure for the leading order terms of a pair of integrals for the system \eqref{Hamiltonian}; however it is still not understood which of them, and under which constraints for the parameters therein, are allowed for non--vanishing magnetic field. (In the absence of linear terms in the Hamiltonian, i.e. natural Hamiltonian, possible forms of the pairs of integrals reduce dramatically, enforcing most of the parameters to vanish \cite{Makarov1967}.) In this paper, we shall search for new integrable systems with the following structure of the leading order terms of the integrals:
\begin{itemize}
	\item generalized elliptic cylindrical:
	\begin{equation}\label{cylell}
		X_1=l_3^2+a l_3 p_3+b p_1^2 +c p_1 p_3+d p_2 p_3+\dots, \quad X_2=p_3^2+\dots, \quad a,b\in\mathbb{R}, b\neq0, c\geq 0, d\geq 0
	\end{equation}
	and $a,c,d$ not all vanishing;
	\item generalized oblate/prolate spheroidal:
	\begin{equation}\label{oblateprolate}
		X_1=l_1^2+l_2^2+l_3^2+a l_3 p_3 + b p_3^2+\dots,\quad X_2=l_3^2+\dots, \quad a,b\in\mathbb{R}\setminus\{0\} ;
	\end{equation}
	\item generalized circular parabolic:
	\begin{equation}\label{circparabolic}
		X_1=l_3^2+\dots,\quad X_2=\frac{1}{2}\left( l_1 p_2+ p_2 l_1 - l_2 p_1-p_1 l_2\right)+a l_3 p_3+\dots,\; a> 0 ;
	\end{equation}
	\item case not directly related to any coordinate system, class (g) in~\cite{Marchesiello2022}:
	\begin{equation}\label{nocoord}
		X_1=l_3^2+a p_3^2+\dots ,\quad X_2= l_3 p_3+b p_3^2+\dots, \quad a,b\in\mathbb{R} .
	\end{equation}
\end{itemize}

\section{The generalized elliptic cylindrical case}\label{sec:cylell}
We present here the solution of the determining equations \eqref{commutH}-\eqref{commutX1X2} for the generalized elliptic cylindrical integrals \eqref{cylell}. 
We assume that at least one of the parameter $a,c,d$ is not zero, otherwise we have the standard elliptic cylindrical type integrals, not generalized ones, and $b\neq0$, as the case $b=0$ was investigated in \cite{KMS2022}.

Let us start by the second order conditions coming from \eqref{commutH}. These read, after we set $s_j^{q_i}=s_j^i$ and $\bm{q}=(x,y,z)$ for simplicity in the notation,

\begin{eqnarray}
	&&(a y-c)B_2(x,y,z)-2 x y B_3(x,y,z)+\partial_{x}s_1^1(x,y,z)=0, \label{eqs21-1} \\
	&&(a x+d)B_1(x,y,z)+2 x y B_3(x,y,z)+\partial_y s_1^2(x,y,z)=0,\label{eqs21-2} \\
	&&-(a x+d)B_1(x,y,z)+(c-a y) B_2(x,y,z)+\partial_z s_1^3(x,y,z)=0, \label{eqs21-3}\\
	&&(c-a y) B_1(x,y,z)-(a x+d) B_2(x,y,z)-2 (b- x^2+ y^2) B_3(x,y,z)\nonumber\\
	&&+\partial_y s_1^1 (x,y,z)+\partial_x s_1^2(x,y,z)=0,\\
	&&(a x +d )B_3(x,y,z)+2 \left(b+y^2\right) B_2(x,y,z)+2 x y B_1(x,y,z)\nonumber\\
	&&+\partial_{z} s_1^{1}(x,y,z)+ \partial_x s_1^3(x,y,z)=0,\\
	&&(a y- c) B_3(x,y,z)-2 x^2 B_1(x,y,z)-2 x y B_2(x,y,z)\nonumber\\
	&&+ \partial_z s_1^2(x,y,z)+\partial_y s_1^3(x,y,z)=0,\label{eqs21-6}
\end{eqnarray}
and
\begin{eqnarray}
	&&\partial_{x}s_2^1(x,y,z)=0,\;\; \partial_{y}s_2^2(x,y,z)=0,\;\; \partial_{z}s_2^3(x,y,z)=0, \label{eqs22-1} \\
	&&\partial_{y}s_2^1(x,y,z)+\partial_{x}s_2^2(x,y,z)=0,\label{eqs22-2}\\ 
	&&2B_1(x,y,z)+\partial_{z}s_2^2(x,y,z)+ \partial_{y}s_2^3(x,y,z)=0 \label{eqB1},\\
	&&2B_2(x,y,z)-\partial_{z}s_2^1(x,y,z)- \partial_{x}s_2^3(x,y,z)=0,\label{eqB2}
\end{eqnarray}
for the integrals $X_1$ and $X_2$, respectively.

Equations \eqref{eqs22-1}-\eqref{eqs22-2} can be easily solved and yield
\begin{equation}\label{solX22S}
	s_2^1(x,y,z)= S_{21}^1(z)+y S_{22}(z),\;\;s_2^2(x,y,z)= S_{21}^2(z)-x S_{22}(z),\;\;s_2^3(x,y,z)=S_2^3(x,y).
\end{equation}
Thus, from \eqref{eqB1}-\eqref{eqB2} we deduce
\begin{eqnarray}
	&&B_1(x,y,z)=-\frac12\left(\partial_{z}S_{21}^2(z) -x \partial_{z} S_{22}(z)+ \partial_{y}S_2^3(x,y)\right)\label{solX22B1},\\
	&&B_2(x,y,z)=\frac12\left(\partial_{z}S_{21}^1(z)+ y \partial_{z} S_{22}(z) +\partial_{x}S_2^3(x,y)\right).\label{solX22B2}
\end{eqnarray}
Instead of solving the remaining second order equations \eqref{eqs21-1}-\eqref{eqs21-6} for the $s_i^j$, it is more convenient to look at the second order conditions coming from \eqref{commutX1X2}, that now simplify due to \eqref{solX22S}-\eqref{solX22B2}. They read
\begin{eqnarray}
	&&2 y S_{21}^2(z)-(c-a y) \left(y\partial_z S_{22}(z)+\partial_z S_{21}^1(z)\right)=0,\label{eqs2-1}\\
	&&(a x+d) \left(x \partial_z S_{22}(z)-\partial_z S_{21}^2(z)\right)+2 x S_{21}^1(z)=0,\label{eqs2-2}\\
	&&2 \partial_z s_1^3(x,y,z)-\left(a \left(x^2+y^2\right)-c y+d x\right)\partial_z S_{22}(z)+(c-a y) \partial_z S_{21}^1\nonumber\\
	&&+(a x+d) \partial_z S_{21}^2(z)=0, \label{eqs2-3}\\
	&&(c x-(2 a x+d) y )\partial_z S_{22}(z)-(a x+ d) \partial_z S_{21}^1(z)+(a y -c)\partial_z S_{21}^2(z)+2 b S_{22}(z)\nonumber\\
	&&-2 x S_{21}^2(z)-2 y S_{21}^1(z)=0,\label{eqs2-4}\\
	&&2 \left(\partial_z s_1^1(x,y,z)+y \left(b+x^2+y^2\right) \partial_z S_{22}(z)+\left(b+y^2\right) \partial_z S_{21}^1(z)-x y \partial_z S_{21}^2(z)\right)\nonumber\\
	&&-a S_{21}^2(z)-d S_{22}(z)=0,\label{eqs2-5}\\
	&&2 \left(\partial_z s_1^2(x,y,z)-x(x^2+y^2) \partial_z S_{22}(z)+x^2 \partial_z S_{21}^2(z)-x y \partial_z S_{21}^1(z)\right)\nonumber\\
	&&+a S_{21}^1(z)+c S_{22}(z)=0.\label{eqs2-6}
\end{eqnarray}
We see that \eqref{eqs2-3} is solved by
\begin{equation}\label{S13sol}
	s_1^3(x,y,z)=S_1^3(x,y)+\frac{1}{2} \left(( a x^2+a y^2-c y+d x)S_{22}(z)+ (a y-c) S_{21}^1(z)- (a x+d) S_{21}^2(z)\right).
\end{equation}
Next we notice that \eqref{eqs2-1},\eqref{eqs2-2} and \eqref{eqs2-4} are polynomials in $x,y$ with coefficients depending on the unknown functions $S_{21}^1(z), S_{21}^2(z), S_{22}(z)$. By equating to zero the coefficients of different monomial terms in $x,y$ we obtain from \eqref{eqs2-1} the condition
\begin{equation}\label{eq:split-a}
	a \partial_z S_{22}(z)=0.
\end{equation}
This leads to two subcases depending on $a\neq0$ or $a=0$ that are solved in the Appendices \ref{sec:anotzero-comp} and \ref{sec:azero-comp}, respectively. For reader's convenience we present only the final results here.

\subsection{$a\neq0$}\label{sec:ellipAnotZero}
If $a\neq0$ we have one new integrable system given by
\begin{equation}\label{BellipAnotZero}
	\vec{B}(x,y,z)=\left(\beta_2 (c-a y), \beta_2 (a x+d) , a\beta_1-3 \beta_2 (x^2+y^2)\right)
\end{equation}
and
\begin{eqnarray}
	W(x,y,z)&=&\beta_2\left[x \left(c d \beta_2 y+ \omega_1\right)-\frac{1}{2} b\beta_2 y^4-\frac{1}{4} \beta_2 x^6-\frac{\beta_2 y^6}{4}+\omega_2 y+\omega_3 y^2\right.\nonumber\\
	& & + x^2 \left(-a b \beta_1 -\frac{1}{2}\left( d^2- c^2\right) \beta_2 -\frac{3}{4} \beta_2 y^4+\omega_3\right) + \frac{\beta_2 x^4}{2} \left(b-\frac{3 y^2}{2}\right)\nonumber\\
	& & +\left.\frac{a}{2}\left(x^2+y^2\right)\left( \frac12\left(\beta_1-\frac{a \beta_2}{2}\right)(x^2+y^2)+ \beta_2( c y- d x)\right)\right],\label{WellipAnotZero}
\end{eqnarray}
where $\beta_i, \omega_i \in\mathbb R$. The functions $s_i^j$, $m_i$ in the integrals are determined by
\begin{eqnarray}
	s_1^1(x,y,z)&=&\frac{1}{2} \beta_2\left(-a^2 y \left(x^2+y^2\right)+a c \left(x^2+3 y^2\right)-2 a d x y-4 b y^3+2 c d x-3 y \left(x^2+y^2\right)^2\right)\nonumber\\
	& & +a \beta_1 y \left(x^2+y^2\right)+\omega_2+2\omega_3 y, \label{s11ellipAnotZero}\\
	s_1^2(x,y,z)&=& \frac{1}{2}\left(x \left(x^2 \left(a^2 \beta_2-2 a \beta_1-4 b \beta_2\right)+4 a b \beta_1+3 a d \beta_2 x-2(c^2- d^2)\beta_2 +3 \beta_2 x^4-4 \omega_3\right)\right.\nonumber\\
	& & - 2\omega_1+ \left.y^2 \left(a x (a \beta_2-2 \beta_1)+a d \beta_2 +6 \beta_2 x^3\right)-2 \beta_2 c y (a x+d)+3\beta_2 x y^4\right),\label{s12ellipAnotZero}\\
	s_1^3(x,y,z)&=&-\frac12\left( a^2 \beta_1 \left(x^2+y^2\right)+ \left(4 b d \beta_2 x+\left(c^2+d^2\right)\beta_1-2 \beta_2 \left(x^2+y^2\right) (d x-c y)\right)\right)\nonumber\\
	& & + \frac{\beta_2}{4a^3}\left(\left(c^2+d^2\right)^2-4 a^2 b d^2\right)+\frac{a}{4} \beta_2 \left(3 \left(x^2+y^2\right)^2-4 b x^2\right)\nonumber\\
	& & +\beta_1 ( c y- d x),\label{s13ellipAnotZero}\\
	m_1(x,y,z)&=&\frac{1}{16 a^3}\left[16 a^2 b \beta_2^2 d^2 (c y-d x)+2 a \beta_2^2 \left(c^2+d^2\right)^2 \left(x^2+y^2\right)-4 \beta_2^2 \left(c^2+d^2\right)^2 (c y-d x)\right.\nonumber\\
	& & - 4 a^6 \beta_1\beta_2\left(x^2+y^2\right)^2-2 a^5 \left[4 b x^2 \left(2\beta_1^2+\beta_2^2 \left(x^2+y^2\right)\right)-\left(x^2+y^2\right) \left(2 \beta_1^2 \left(x^2+y^2\right)\right.\right.\nonumber\\
	& & +\left.\left. 8 \beta_1\beta_2(c y-d x)+3 \beta_2^2 \left(x^2+y^2\right)^2\right)\right]
	+4a^4\left[4 \beta_1 y (2 \beta_2 c d x+\omega_2)\right.\nonumber\\
	& & +\beta_1\left(x \left(\beta_2 x \left(-8 b^2+10 b x^2+c^2-5 d^2-3 x^4\right)+4 \omega_1+4\omega_3 x\right)\right.\nonumber\\
	& & - \left.y^2 \left(\beta_2 \left(-8 b x^2+3 c^2+d^2+9 x^4\right)-4\omega_3\right)-\beta_2 y^4 \left(2 b+9 x^2\right)-3 \beta_2 y^6\right)\nonumber\\
	& & +\left.\beta_2^2 \left(5 \left(x^2+y^2\right)^2 (d x-c y)-4 b x \left(-c x y+2 d x^2+d y^2\right)\right)\right]\nonumber\\
	& & +a^3\beta_2\left[16 b^2 \beta_2 x^4+8 \beta_1 \left(c^2+d^2\right) (c y-d x)+(x^2+y^2)\left(-16 \omega_1 x-8 \left(2 \omega_2 y+3\omega_3 \left(x^2+y^2\right)\right)\right.\right.\nonumber\\
	& & +\left.\beta_2\left(4 c^2 \left(y^2-3 x^2\right)-32 c d x y+16 d^2 x^2+9 \left(x^2+y^2\right)^3\right)\right)+32 b x (\omega_1+ \omega_3 x)\nonumber\\
	& & +\left.\left.8 b \left(\beta_2 \left(2 c^2 x^2+4 c d x y-d^2 \left(5 x^2+y^2\right)-\left(3 x^2-2 y^2\right) \left(x^2+y^2\right)^2\right)\right)\right]\right],	\label{m1ellipAnotZero}
\end{eqnarray}
and
\begin{equation}\label{s2ellipAnotZero}
	s_2^1(x,y,z)=s_2^2(x,y,z)=0,\; s_2^3(x,y,z)=\frac{\beta_2 \left((c-a y)^2+(a x+d)^2\right)}{a},
\end{equation}
\begin{equation}\label{m2ellipAnotZero}
	m_2(x,y,z)=\frac{\beta_2^2 \left((c-a y)^2+(a x+d)^2\right)^2}{4 a^2}
\end{equation}
for the integrals $X_1$ and $X_2$, respectively. Thus, if the gauge is chosen so that $A_3=-s_2^3$, $X_2=p_3$.

Also notice that for $b=0$ the integral $X_1$ \eqref{oblateprolate} reduces to generalized cylindrical type integral. Up to translation in $x$ we can then simplify to $d=0$ and indeed the system \eqref{BellipAnotZero}-\eqref{WellipAnotZero} reduces to system (3.32) in \cite{KMS2022}.

\subsection{$a=0$}\label{sec:ellipAZero}
If $c d\neq0$ we just obtain a subcase of system \eqref{BellipAnotZero}, \eqref{WellipAnotZero} for $a=0$.

We find new integrable system if $c$ or $d$ are equal to zero (not both, otherwise we have standard elliptic cylindrical integral). Without loss of generality, we consider only the case $d=0$. Only the final results are provided here for reader convenience. If $c=0$, we apply the permutation $(x,y,p_1,p_2)\to (y,x,p_2,p_1)$ and consider the integral $\tilde X_1=X_1-b (2 H- X_2)$. Since under the permutation $\ell_3\to-\ell_3$ and $p_3$ does not change,
$$\tilde X_1=\ell_3^2- a l_3 p_3+ d p_1 p_3- b p_1^2+\dots, $$
and we reduce to the case $d=0$ (up to renaming the constant $d$ as $c$ and a change in the sign of $b$).

The form of the solution depends on $\epsilon=\text{sgn} b $. When $b<0$ the potential and the magnetic field are periodic in the $z$ coordinate, when $b>0$ they exponentially blow up with $z\rightarrow \pm \infty$. 

The magnetic field and the potential read
\begin{equation}\label{mag_ell_a0}
	\vec{B}(x,y,z)=\left(\frac{\delta}{4} \left( 2 x \mathcal U_1(z)+ c \delta\mathcal U_2(z) \right)-c \beta_2,\; \frac{\delta }{2} y \mathcal U_1(z),\; 3 \beta_2 \left(x^2+y^2\right)+\epsilon \mathcal U_2(z) -\beta_1\right)
\end{equation}
and
\begin{eqnarray}
	W(x,y,z)&=&-\epsilon \frac{c^2 \delta^4+4 \left( \epsilon \delta^2( x^2+y^2)-1\right)}{64 \delta^2}(\mathcal U_2(z)^2+\epsilon\mathcal U_1(z)^2)
	-\epsilon \frac{\delta c x }{8} \mathcal U_1(z) \mathcal U_2(z)\nonumber\\
	& &
	-\epsilon \frac{c^2 x \left( -\epsilon \beta_1 \delta^2+\beta_2 \left(\epsilon\delta^2( x^2+ y^2)+2\right)\right)+2 \omega_1}{4 c \delta }\mathcal U_1(z)
	\nonumber\\
	&& -\frac{\epsilon}{8} \left(\epsilon \beta_2 c^2 \delta^2 y^2+\left(x^2+y^2\right) \left(3 \beta_2\left(x^2+y^2\right)-2 \beta_1\right)+2 \omega_2\right) \mathcal U_2(z)
	\nonumber\\
	& &+\frac{1}{16}\left[-4 \beta_2^2 x^6+2 \beta_2 x^4 \left(2 \beta_1+\epsilon\beta_2 c^2 \delta^2-6 \beta_2 y^2\right)+16\beta_2 \omega_1 x\right.\nonumber\\
	&& -y^2 \left(\alpha_1^2-\epsilon \alpha_2^2+8\beta_2 \omega_2+2 \beta_2 y^2 \left(-2 \beta_1+\epsilon c^2 \delta^2 \beta_2+2\beta_2 y^2\right)\right)\nonumber\\
	&& -\left. x^2 \left(\alpha_1^2-\epsilon \alpha_2^2+4\beta_2 \left(c^2 \left(\epsilon\beta_1 \delta^2-2 \beta_2\right)+3 \beta_2 y^4-2 \beta_1 y^2+2 \omega_2\right)\right)\right]
	\label{pot_ell_a0}
\end{eqnarray}
respectively, where 
\begin{equation}
	\mathcal U_1(z)=\alpha_1\sinh(\delta z)+\alpha_2\cosh(\delta z), \quad 
	\mathcal U_2(z) =-\alpha_2\sinh(\delta z)-\alpha_1\cosh(\delta z), \quad \delta=\frac{2}{c}\sqrt{b},
\end{equation} 
for $b>0$, i.e. $\epsilon=1$, and 
\begin{equation}\label{U1U2}
	\mathcal U_1(z)=\alpha_2\cos(\delta z)-\alpha_1\sin(\delta z), \quad \mathcal U_2(z)=\alpha_2\sin(\delta z)+\alpha_1\cos(\delta z), \quad \delta=\frac{2}{c}\sqrt{-b},
\end{equation}
for $b<0$, i.e. $\epsilon=-1$. In both cases we have a relation between the functions $\mathcal U_1(z)$ and $\mathcal U_2(z)$ of the form $\mathcal U_2(z)=-\frac{1}{\delta} \mathcal U_1'(z)$.

We notice that the solutions for $b$ negative and positive can be viewed as related by an analytic continuation $\delta\rightarrow \mathfrak{i} \delta$, with appropriate redefinition of the parameters $\alpha_1$ and $\alpha_2$ to have explicitly real solutions. 

The functions $s_i^j$, $m_i$ in the integrals are given by
\begin{eqnarray}
	s_1^1(x,y,z)&=&\frac{1}{4} y \left[2 c \delta x \mathcal U_1(z)+\left(c^2 \delta^2+4 \epsilon (x^2+y^2)\right)\mathcal U_2(z)+2 \beta_2 (\epsilon c^2 \delta^2 y^2+3 x^4) \right.\nonumber\\
	& & - \left. 4 x^2 \left(\beta_1-3 \beta_2 y^2\right)+6 \beta_2 y^4-4 \beta_1 y^2+4 \omega_2 \right],\nonumber\\
	s_1^2(x,y,z)&=&\frac{1}{2\delta}\left[c^2 \left(\epsilon \beta_2\delta^3 x^3-\epsilon \beta_1 \delta^3 x+2 \beta_2 \delta x\right)-c \left(\epsilon+\delta^2 x^2\right)\mathcal U_1(z)+\delta \left(-2\epsilon x \left(x^2+y^2\right)\mathcal U_2(z)\right.\right.\nonumber\\
	& & - \left.\left. 3 \beta_2 x^5+2 x^3 \left(\beta_1-3 \beta_2 y^2\right)+x \left(-3 \beta_2 y^4+2\beta_1 y^2-2 \omega_2\right)+2\omega_1\right)\right],\nonumber\\
	s_1^3(x,y,z)&=&\frac{1}{2} c y \left(-2 \beta_1+\epsilon\mathcal U_2(z)+2 \beta_2 \left(x^2+y^2\right)\right),
\end{eqnarray}

\begin{eqnarray}
	m_1(x,y,z)&=& 	\frac{1}{16 (\epsilon \alpha_2^2 - \alpha_1^2) \delta}\left[\delta \left(4 (x^2 + y^2)^2 + \epsilon c^2 \delta^2 (x^2 + y^2) + 2 c^2\right) \left(\alpha_2 \mathcal U_1(z) - \alpha_1 \mathcal U_2(z) \right) \right. \nonumber \\
	& & + \left. 4 c x \left(\epsilon \delta^2 (x^2 + y^2) + 1\right) \left(\epsilon \alpha_2 \mathcal U_2(z)-\alpha_1 \mathcal U_1(z) \right)\right] \left(\alpha_1 \mathcal U_2(z) + \alpha_2 \mathcal U_1(z)\right)
	\nonumber\\
	& & - \frac{c}{8 \delta} \left[-3 \delta^2 \beta_{2} x^5+\left(\epsilon \beta_{2} \left( c^2 \delta^{4} -4 \right) +\delta^2\left(2 \beta_{1}-6 \beta_{2} y^2\right) \right) x^{3}\right.\nonumber\\
	& & +\left.
	\left( \delta^2 \left(\left(2 c^2-3 y^{4}\right) \beta_{2}+2 \beta_{1} y^2-2 \omega_2\right) - \epsilon \left( \beta_{1} \left(c^2 \delta^4 -4 \right) +4 \beta_{2} y^2 \right)\right) x +2 \delta^2 \omega_1\right]
	\mathcal U_1(z) \nonumber\\
	& & +\frac{1}{4}\left[3 \epsilon \beta_{2} x^6+\left(\beta_{2}\left(9 \epsilon y^2-c^2 \delta^2\right) - 2\epsilon \beta_{1}\right) x^{4}+\left(\epsilon \left(\left(9 y^{4}-c^2\right) \beta_{2}-4 \beta_{1} y^2+2 \omega_2\right) +c^2 \delta^2 \beta_{1}\right) x^2\right.\nonumber\\
	& & - \left.2 \epsilon \omega_1 x +\epsilon \left(3 \beta_{2} y^6-2 \beta_{1} y^4+\left(c^2 \beta_{2}+2 \omega_2\right) y^2-c^2 \beta_{1}\right) +c^2 \delta^2 \beta_{2} y^4 \right]\mathcal U_2(z)\nonumber\\
	& & +\frac{9 \beta_2^2}{16}( x^8+ y^8)+\frac{1}{8}\left(18 \beta_{2}^2 y^2-3 \beta_{2} \left(c^2 \delta^2 \epsilon \beta_{2}+2 \beta_{1}\right) \right) x^6\nonumber\\
	& & +\frac{1}{16}\left[54 \beta_{2}^2 y^{4}- \beta_{2} \left( 8 c^2 \delta^2 \epsilon \beta_{2}+36 \beta_{1}\right) y^2+c^2 \left(c^{2} \delta^{4} - 12 \right) \beta_{2}^2 \right. \nonumber \\ & + & \left. \left(10 \epsilon c^2 \delta^2 \beta_{1}+12 \omega_2\right) \beta_{2}-4 \epsilon \alpha_{2}^2+4 \beta_{1}^2\right] x^{4}\nonumber\\
	& & +\frac{1}{4} \left(c \delta \alpha_{1} \alpha_{2}- 4 \beta_{2} \omega_1\right)x (x^2+y^2)+\frac{1}{16}\left[36 \beta_{2}^2 y^6+2 \beta_{2} \left(\epsilon c^2 \delta^2 \beta_{2}-18 \beta_{1}\right) y^{4}\right.\nonumber\\
	& & +\left.\left( \left(8 \epsilon c^2 \delta^2 \beta_{1}- 8 c^2 \beta_{2}+ 24 \omega_2\right) \beta_{2}-8\epsilon \alpha_{2}^2+8\beta_{1}^2\right) y^2-2 \delta^2 c^4 \beta_{2} \left( \delta^{2} \beta_{1} -2\epsilon \beta_{2}\right) \right.\nonumber \\
	& & +\left. c^2 \left(8 \beta_{1} \beta_{2}- \epsilon \,\delta^2 \left(4 \omega_2 \beta_{2}+4 \beta_{1}^2+\alpha_{1}^2\right) \right)-8 \omega_2 \beta_{1}\right] x^2\nonumber\\
	& & +\frac{1}{4 \delta}\left( \epsilon \left( 2 c^2 \delta^{3} \beta_{2} \omega_1 + c \alpha_{1} \alpha_{2}\right) +4 \delta \beta_{1} \omega_1 \right) x +\frac{\beta_{2}}{4} \left( \epsilon c^2 \delta^2 \beta_{2}-3 \beta_{1}\right) y^6\nonumber\\
	& & +\frac{1}{8}\left(2 \beta_{1}^2+2 c^2 \beta_{2}^2+\left(6 \omega_2 - \epsilon c^2 \delta^2 \beta_{1}\right) \beta_{2}-2 \epsilon \alpha_{2}^2\right) y^{4}\nonumber\\
	& & - \frac{1}{16}\left( 8 \left(c^2 \beta_{2}+\omega_2\right) \beta_{1}+ c^2 \delta^2 \alpha_{2}^2\right) y^2
\end{eqnarray}
and
\begin{equation}
	s_2^1(x,y,z)= -\epsilon y\mathcal U_2(z),\;\;s_2^2(x,y,z)=\frac{c \delta \mathcal U_1(z)}{2}+\epsilon x \mathcal U_2(z), \;\;s_2^3(x,y,z)=2 \beta_2 c y,
\end{equation}
\begin{eqnarray}
	m_2(x,y,z)&=&\frac{1}{16} \left(-c^2 \delta^2+\frac{2}{\delta^2}-4 \epsilon (x^2+y^2)\right)\left(\mathcal U_1(z)^2+\epsilon\mathcal U_2(z)^2\right)-\epsilon \frac{c \delta x}{2} \mathcal U_1(z)\mathcal U_2(z)\nonumber\\
	& & -  \left(\epsilon \frac{\omega_1}{c \delta }+\frac{1}{2} c \delta x \left(\beta_2 \left(x^2+y^2\right)-\beta_1\right)\right)\mathcal U_1(z)\nonumber\\
	& & -  \frac{\epsilon}{4} \left(-2 \beta_2 c^2-y^2 \left(2\beta_1-\epsilon c^2 \delta^2\beta_2\right)+3 \beta_2 x^4-2 x^2 \left(\beta_1-3 \beta_2 y^2\right)+3 \beta_2 y^4+2 \omega_2\right)\mathcal U_2(z)\nonumber\\
	& & + c^2 \beta_2^2 y^2+ \frac{1}{4}(\epsilon \alpha_2^2- \alpha_1^2 ) (x^2 + y^2),
\end{eqnarray}
respectively.

\begin{figure}
\resizebox{0.5\hsize}{!}{\includegraphics*{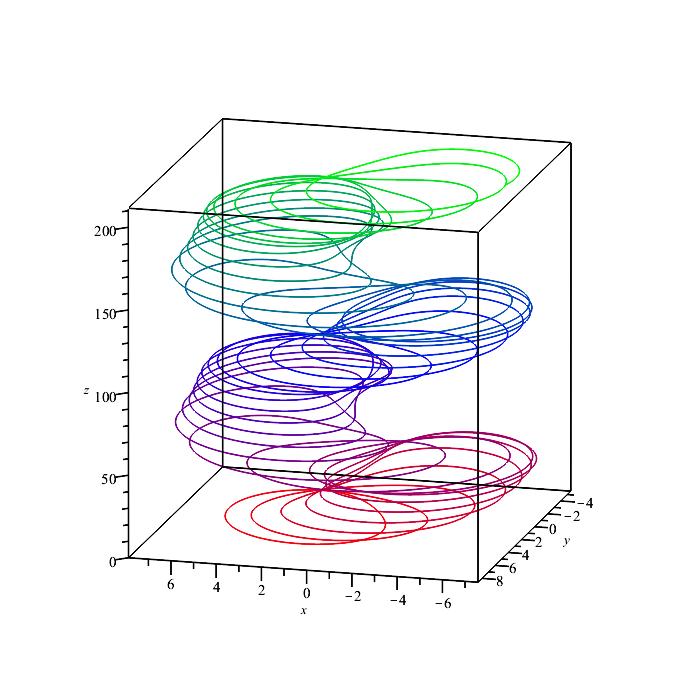}}\resizebox{0.5\hsize}{!}{\includegraphics*{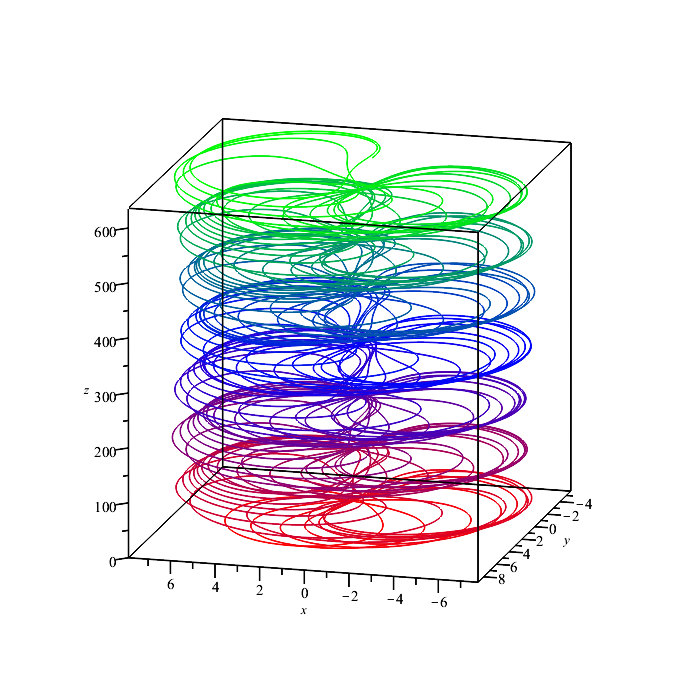}}
\caption{Trajectories of the system \eqref{mag_ell_a0}--\eqref{pot_ell_a0} for the parameters~\eqref{picparams}, the initial conditions~\eqref{picchoice1} and the time $t\in(0,50)$ (left plot) and $t\in(0,150)$ (right plot).}
\label{pic1} 
\end{figure}

\begin{figure}
\resizebox{0.5\hsize}{!}{\includegraphics*{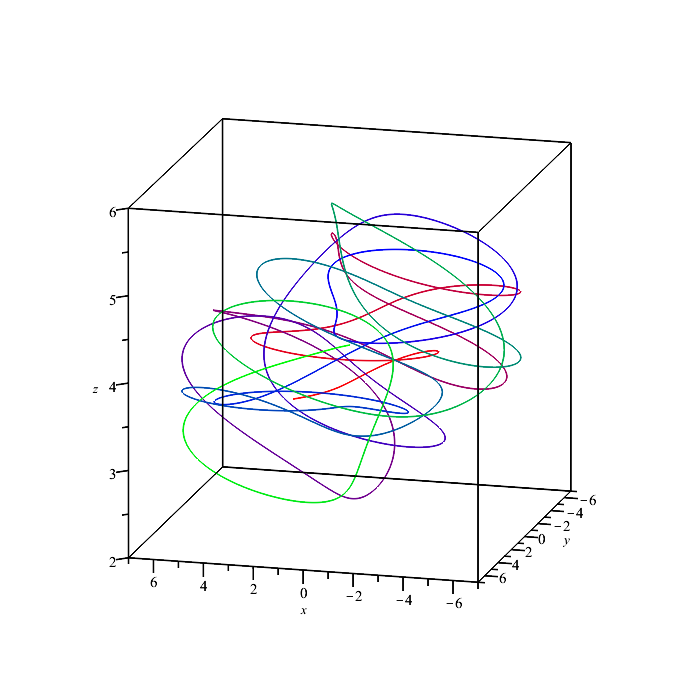}}\resizebox{0.5\hsize}{!}{\includegraphics*{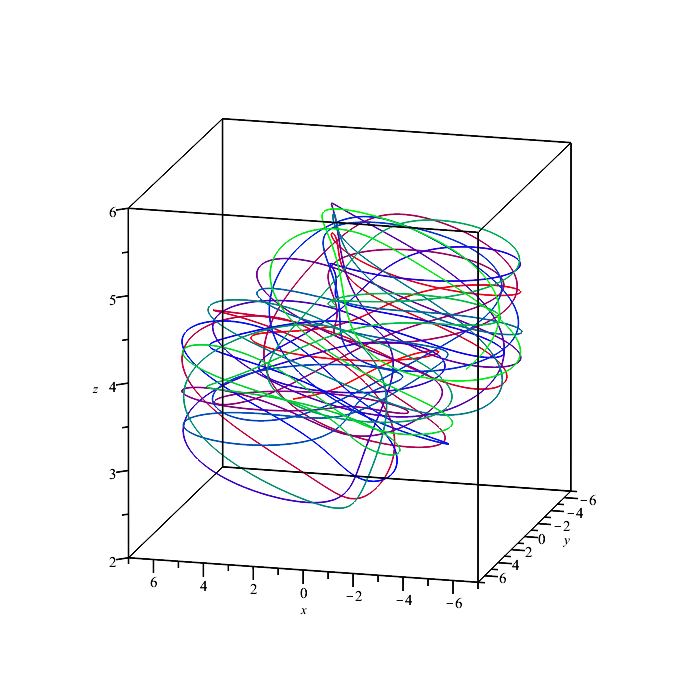}}
\caption{Trajectories of the system \eqref{mag_ell_a0}--\eqref{pot_ell_a0} for the parameters~\eqref{picparams}, the initial conditions~\eqref{picchoice2} and the time $t\in(0,20)$ (left plot) and $t\in(0,50)$ (right plot).}
\label{pic2} 
\end{figure}

As the magnetic field and the electrostatic potential are periodic in the $z$ coordinate when $b<0$ and thus appear to be of most interest for applications, we tried to obtain some intuition about the behaviour of particles in these fields by numerically constructing sample trajectories for generic values of the system's parameters. Namely we chose the parameters as
\begin{equation}\label{picparams}
\delta = 1, \, c = 2, \, \alpha_1 = 1, \, \alpha_2 = 1, \, \beta_1 = -\frac{1}{5}, \, \beta_2 =-\frac{1}{7}, \, \omega_1 = -1, \, \omega_2 = -1
\end{equation}
and plotted trajectories for two choices of the initial conditions
\begin{equation}\label{picchoice1}
x(0) = \pi, \, y(0) = -\pi, \, z(0) = 0, \, p_1(0) = 1, \, p_2(0) = 0, \, p_3(0) = 1
\end{equation}
and
\begin{equation}\label{picchoice2}
x(0) = \pi, \, y(0) = -\pi, \, z(0) = \frac{\pi}{2}, \, p_1(0) = 1, \, p_2(0) = 0, \, p_3(0) = 1.
\end{equation}
For the first choice~\eqref{picchoice1}, the trajectories are clearly unbounded in the $z$--direction, whereas for the second choice~\eqref{picchoice2} they appear to be bounded, see Figs~\ref{pic1} and~\ref{pic2} and the indicated ranges for the spatial axes for the two ranges of the time parameter.

\section{Other cases}\label{sec:others}

\subsection{The generalized prolate / oblate spheroidal case}
\label{sec:spher}

Let us now focus on the generalized prolate / oblate spheroidal integrals \eqref{oblateprolate}. We assume that both of the parameters $a,b$ are not zero. If $b=0$ the pair of integrals reduces to the generalized spherical case investigated in~\cite{KMS2022} (establishing that no new system can be found of this form), whereas if $a=0$ we have standard form of prolate / oblate spheroidal integrals and such systems were classified in~\cite{BertrandSnobl}. In this case, it is more convenient to rewrite our integrals and equations determining them in the cylindrical coordinates
\newcommand{\cphi}{\varphi}
\begin{equation}\label{cyl coords}
	x=r \cos\cphi,\quad y=r \sin\cphi,\quad z=Z.
\end{equation}
The general structure of the Hamiltonian~\eqref{Hamiltonian}, the integrals and the determining equations become as in section 2 of~\cite{Fournier2019} where we refer  for further details.

For our choice of the structure of the commuting pair of integrals~\eqref{oblateprolate}, the leading order determining equations coming from \eqref{commutH}-\eqref{commutX1X2} in cylindrical coordinates read
\begin{eqnarray}\label{oblateprolateHX1}
	\nonumber & & 2 r Z B_\cphi(r, \cphi, Z) + a B_r(r, \cphi, Z)-\partial_Z s^Z_1(r, \cphi, Z)=0, \\
	\nonumber & & 2 r Z B_Z(r, \cphi, Z)+ 2 (b - Z^2) B_r(r, \cphi, Z) + r^2 \partial_Z s^\cphi_1(r, \cphi, Z) +\partial_\cphi s^Z_1(r, \cphi, Z)=0, \\ \nonumber
	& & 2 ( Z^2 - r^2 - b ) B_\cphi(r, \cphi, Z) + a B_Z(r, \cphi, Z) + \partial_Z s^r_1(r, \cphi, Z)+\partial_r s^Z_1(r, \cphi, Z)=0, \\
	& & a r B_r(r, \cphi, Z) +r \partial_\cphi s^\cphi_1(r, \cphi, Z)+s^r_1(r, \cphi, Z) =0,\\
	\nonumber & & 2 r^2 B_Z(r, \cphi, Z) -a r^2 B_\cphi(r, \cphi, Z) - 2 r Z B_r(r, \cphi, Z) + r^2 \partial_r s^\cphi_1(r, \cphi, Z)+\partial_\cphi s^r_1(r, \cphi, Z)=0, \\
	\nonumber & & 2 r Z B_\cphi(r, \cphi, Z)+ \partial_r s^r_1(r, \cphi, Z) =0
\end{eqnarray}
from $\{H,X_1 \}=0$,
\begin{eqnarray}\label{oblateprolateHX2}
	\nonumber & & \partial_Z s^Z_2(r, \cphi, Z) = 0, \\
	\nonumber & & r^2 \left(-2 B_r(r, \cphi, Z) + \partial_Z s^\cphi_2(r, \cphi, Z) \right)+ \partial_\cphi s^Z_2(r, \cphi, Z) =0, \\
	& & \partial_Z s^r_2(r, \cphi, Z)+\partial_r s^Z_2(r, \cphi, Z)=0, \\
	\nonumber & & \frac{1}{r} s^r_2(r, \cphi, Z)+\partial_\cphi s^\cphi_2(r, \cphi, Z)=0, \\
	\nonumber & & r^2 \left(2 B_Z(r, \cphi, Z) +\partial_r s^\cphi_2(r, \cphi, Z) \right)+\partial_\cphi s^r_2(r, \cphi, Z)=0, \\
	\nonumber & & \partial_r s^r_2(r, \cphi, Z) = 0
\end{eqnarray}
from $\{H,X_2 \}=0$, and finally
\begin{eqnarray}\label{oblateprolateX1X2}
	\nonumber && - a r^2 \partial_Z s^\cphi_2(r, \cphi, Z) + 2 r \left( s^r_2(r, \cphi, Z) - s^r_1(r, \cphi, Z) \right) + 2 Z s^Z_2(r, \cphi, Z) = 0, \\
	\nonumber && 2 r (Z^2 - b) \partial_Z s^\cphi_2(r, \cphi, Z) + 2 Z r^2 \partial_r s^\cphi_2(r, \cphi, Z) - 2 r^3 \partial_Z s^\cphi_1(r, \cphi, Z) + a s^r_2(r, \cphi, Z) = 0,
	\\ 
	&& r \partial_Z s^r_2(r, \cphi, Z) + s^Z_2(r, \cphi, Z) = 0, \\
	\nonumber && a r \left( 2 B_r(r, \cphi, Z) - \partial_Z s^\cphi_2(r, \cphi, Z) \right) + 2 Z \partial_Z s^r_2(r, \cphi, Z) - 2 s^r_2(r, \cphi, Z) = 0, \\ \nonumber
	&& 2 ( r^2 - Z^2+ b) \partial_Z s^r_2(r, \cphi, Z) + a \partial_\cphi s^r_2(r, \cphi, Z) + 2 Z s^r_2(r, \cphi, Z) + 2 r s^Z_2(r, \cphi, Z) = 0,
	\\ \nonumber && 2 r^2 \left( a B_\cphi(r, \cphi, Z) + \partial_r s^\cphi_2(r, \cphi, Z) - \partial_r s^\cphi_1(r, \cphi, Z) \right) + 2 r Z \partial_Z s^\cphi_2(r, \cphi, Z) - \\ \nonumber && - a \partial_Z s^r_2(r, \cphi, Z) = 0
\end{eqnarray}
from $\{X_1,X_2 \}=0$ (where equations~\eqref{oblateprolateHX1} and~\eqref{oblateprolateHX2} were used for simplification).

The equations~\eqref{oblateprolateHX2} determining the second integral were encountered before, e.g. in.~\cite{Marchesiello2018Sph,Fournier2019,KMS2022}, and their solution is known. It reads
\begin{align}\label{Lz2_int_sol}
	B_Z(r, \cphi, Z) & = -\frac{1}{2 r^2}\left( r^2 \partial_r S_2^\cphi(r, Z) + Z \left( \partial_{\cphi \cphi} S^r_{21}(\cphi)+S^r_{21}(\cphi) \right) +\partial_{\cphi \cphi} S^r_{20}(\cphi) + S^r_{20}(\cphi) \right), \nonumber \\
	B_r(r, \cphi, Z) & = \frac{1}{2 r^2}\left( r^2 \partial_Z S_2^\cphi(r, Z) - r \left( \partial_{\cphi \cphi} S^r_{21}(\cphi)+S^r_{21}(\cphi) \right)+ \partial_\cphi S^Z_{20}(\cphi)\right), \nonumber \\
	s_2^r(r, \cphi, Z) & = Z \partial_\cphi S^r_{21}(\cphi) + \partial_\cphi S^r_{20}(\cphi), \nonumber \\
	s_2^\cphi(r, \cphi, Z) & = -\frac{1}{r} \left(Z S^r_{21}(\cphi) + S^r_{20}(\cphi)\right) + S_2^\cphi(r, Z), \nonumber \\
	s_2^Z(r, \cphi, Z) & = - r \partial_\cphi S^r_{21}(\cphi) + S^Z_{20}(\cphi),
\end{align}
where newly introduced functions $S^{\cdot}_{\ldots}(\ldots)$ are arbitrary functions of their respective variables.

The commutativity conditions~\eqref{oblateprolateX1X2} imply that the function $S^Z_{20}(\cphi)$ must identically vanish, that function $S^r_{21}(\cphi)$ determines $S^r_{20}(\cphi)$ up to an arbitrary constant,
\begin{equation}
	S^r_{20}(\cphi) = -\frac{a}{2}\frac{{\mathrm d} S^r_{21}(\cphi)}{{\mathrm d} \cphi} + S^r_{200},
\end{equation}
and that function $S^r_{21}(\cphi)$ solves a third order ODE
\begin{equation}
	\frac{{\mathrm d^3} S^r_{21}(\cphi)}{{\mathrm d} \cphi^3}-\frac{4 b}{a^2} \frac{{\mathrm d} S^r_{21}(\cphi)}{{\mathrm d} \cphi} =0,
\end{equation}
thus 
\begin{equation}
	S^r_{21}(\cphi) = S^r_{211} + S^r_{212} {\mathrm{e}}^{2 \frac{\sqrt{b}}{a} \cphi} + S^r_{213} {\mathrm e}^{-2 \frac{\sqrt{b}}{a} \cphi}
\end{equation}
or	
\begin{equation}
	S^r_{21}(\cphi) = S^r_{211} + S^r_{212} \cos\left( 2 \frac{\sqrt{|b|}}{a} \cphi \right)+ S^r_{213} \sin\left( 2 \frac{\sqrt{|b|}}{a} \cphi \right),
\end{equation}
depending on the sign of the parameter $b$, i.e. whether we consider generalized prolate or oblate spheroidal case. The remaining equations in~\eqref{oblateprolateX1X2} determine the dependence of the functions $s_1^{r,\cphi,Z}(r, \cphi, Z)$ on the angular variable $\cphi$,
\begin{align}
	s_1^Z(r, \cphi, Z) = & s_1^Z(r, Z)+ \frac{ a^2 + 4 b}{2 a r} \left( \sqrt{b} \left( S^r_{212} {\mathrm{e}}^{2 \frac{\sqrt{b}}{a} \cphi} - S^r_{213} {\mathrm{e}}^{-2 \frac{\sqrt{b}}{a} \cphi} \right) -
	\left( S^r_{212} {\mathrm{e}}^{2 \frac{\sqrt{b}}{a} \cphi}+ S^r_{213} {\mathrm{e}}^{-2 \frac{\sqrt{b}}{a} \cphi}\right) Z \right)
	, \nonumber \\ 
	s_1^\cphi(r, \cphi, Z) = & s_1^\cphi(r, Z) + 2 \frac{\sqrt{b}}{r} \left(S^r_{212} {\mathrm{e}}^{2 \frac{\sqrt{b}}{a} \cphi}- S^r_{213} {\mathrm{e}}^{-2 \frac{\sqrt{b}}{a} \cphi} \right), \\
	s_1^r(r, \cphi, Z) = & s_1^r(r, Z)+ \frac{a^2 - 4 b}{2 a}\left( S^r_{212} {\mathrm{e}}^{2 \frac{\sqrt{b}}{a} \cphi} +S^r_{213} {\mathrm{e}}^{-2 \frac{\sqrt{b}}{a} \cphi} \right) , \nonumber 
\end{align}
and similarly in terms of trigonometric functions $\cos\left( 2 \frac{\sqrt{|b|}}{a} \cphi \right)$ and $\sin\left( 2 \frac{\sqrt{|b|}}{a} \cphi \right)$ for $b$ negative.

Next, the equations for the first integral~\eqref{oblateprolateHX1} fix
the last component of the magnetic field 
\begin{equation}
	B_\cphi(r, \cphi, Z) = \frac{2 r \partial_Z S_1^Z(r, Z)-a r \partial_Z S_2^\cphi(r, Z) + a S^r_{211}}{4 r^2 Z }.
\end{equation}
The remaining equations in~\eqref{oblateprolateHX1} determine the first order derivatives $\partial_Z S_1^Z(r, Z)$, $\partial_r S_1^Z(r, Z)$, $\partial_Z S_1^\cphi(r, Z)$ and $\partial_r S_1^\cphi(r, Z)$
in terms of the constants $S^r_{200}, S^r_{211}$, functions $S_2^\cphi(r, Z)$ and $S_1^r(r, Z)$ and their derivatives, together with the last equation
\begin{equation}\label{oblateprolateHX1last}
	a r \partial_Z S_2^\cphi(r, Z) = a S^r_{211} - 2 S_1^r(r, Z).
\end{equation}	
Equating the expressions obtained for second order mixed derivatives from the equations determining $\partial_Z S_1^Z(r, Z)$, $\partial_r S_1^Z(r, Z)$, $\partial_Z S_1^\cphi(r, Z)$ and $\partial_r S_1^\cphi(r, Z)$, i.e. 
\begin{equation}\label{comp_conds}
	\partial_r\left(\partial_Z S_1^Z(r, Z)\right)=\partial_Z\left(\partial_r S_1^Z(r, Z)	\right), \quad \partial_r\left(\partial_Z S_1^\cphi(r, Z)\right)=\partial_Z\left(\partial_r S_1^\cphi(r, Z)\right),
\end{equation}
and combining them with~\eqref{oblateprolateHX1last} we arrive at a system of PDEs for two functions $S_2^\cphi(r, Z)$ and $S_1^r(r, Z)$. This system implies further compatibility conditions for the derivatives of the functions $S_2^\cphi(r, Z)$ and $S_1^r(r, Z)$ and using them it can be solved, introducing several more integration constants and fixing any remaining functional dependence in $s_1^{r,\cphi,Z}(r, \cphi, Z)$ and $s_2^{r,\cphi,Z}(r, \cphi, Z)$. For generic values of the parameters $a,b$ the complete solution of the equations~\eqref{oblateprolateHX1}--\eqref{oblateprolateX1X2} reads
\begin{eqnarray}\label{oblateprolate_leading_order_solved}
	B_r(r, \cphi, Z) & = &-\frac{1}{2 a^2 r^4}\left[\left(a^2 + 4 b\right) r^3 \left( S^r_{212} {\mathrm e}^{2 \frac{\sqrt{b}}{a} \cphi}+ S^r_{213}{\mathrm e}^{-2 \frac{\sqrt{b}}{a} \cphi} \right) + 2 a Z \left(S^r_{110} r^2 + S^r_{120} \left(Z^2 - b\right)\right) \right], 
	\nonumber \\
	B_\cphi(r, \cphi, Z) & = & \frac{1}{2 r^5}\left(S^r_{120} \left(3 Z^2 - 3 b\right) + S^r_{110} r^2 \right), \nonumber \\
	B_Z(r, \cphi, Z) & = & -\frac{1}{8 a^2 r^5} \left[4 \left(a^2 + 4 b\right) r^3 \left( S^r_{212} \left(Z-\sqrt{b} \right) {\mathrm e}^{2 \frac{\sqrt{b}}{a} \cphi} + \left(Z+ \sqrt{b} \right) S^r_{213} {\mathrm e}^{-2 \frac{\sqrt{b}}{a} \cphi} \right) \right. + \nonumber \\ & & \left.+ 8 a \left(\frac{a}{2} S^\cphi_{211} r^6+ \left( S^r_{110} Z^2-\frac{3 a^2}{8} S^r_{120} - b \left(S^r_{110} - \frac{S^r_{120}}{2}\right)\right) r^2 + S^r_{120} \left(Z^2 - b\right)^2\right) \right], \nonumber \\
	s_1^r(r, \cphi, Z) & =& \frac{1}{2 a r^3} \left[\left(a^2 - 4 b\right) r^3 \left( S^r_{212} {\mathrm e}^{2 \frac{\sqrt{b}}{a} \cphi} +S^r_{213} {\mathrm e}^{-2 \frac{\sqrt{b}}{a} \cphi} \right) + 2 a Z \left(S^r_{110} r^2 + S^r_{120} \left(Z^2 - b\right)\right) \right], \nonumber \\
	s_1^\cphi(r, \cphi, Z) & = & \frac{1}{8 a r^4 } \left[ 16 a \sqrt{b} r^3 \left( S^r_{212} {\mathrm e}^{2 \frac{\sqrt{b}}{a} \cphi} - S^r_{213} {\mathrm e}^{-2 \frac{\sqrt{b}}{a} \cphi} \right) + \right. \nonumber \\ & & \left. +a^2 \left(\left(-2 S^r_{110} + 3 S^r_{120}\right) r^2 - 3 S^r_{120} \left(Z^2 - b\right)\right) + 4 a r^4 \left(Z^2 S^\cphi_{211} + S^\cphi_{211} r^2+ 2 S^\cphi_{10}\right) \right. + \nonumber \\ & & \left. + 4 b \left(\left(2 S^r_{110} - S^r_{120}\right) r^2 + S^r_{120} \left(Z^2 - b\right)\right)\right],
	\\
	s_1^Z(r, \cphi, Z) & = & \frac{1}{16 a r^4} \left[- 8 \left(a^2 + 4 b\right) r^3\left( S^r_{212} \left(Z-\sqrt{b} \right) {\mathrm e}^{2 \frac{\sqrt{b}}{a} \cphi}+ S^r_{213}\left(Z+ \sqrt{b} \right) {\mathrm e}^{-2 \frac{\sqrt{b}}{a} \cphi} \right) \right. + \nonumber \\ & & \left.+ 8 a \left(\frac{a}{2} S^\cphi_{211} r^6+ 2 S^Z_{10} r^4 + \left(\frac{3}{8} S^r_{120} \left(a^2 + 4 b\right) r^2\right) + S^r_{120} \left(Z^2 - b\right)^2\right)\right], \nonumber \\
	s_2^r(r, \cphi, Z) & = & \frac{2 \sqrt{b}}{a} \left( S^r_{212} \left(Z - \sqrt{b}\right) {\mathrm e}^{2 \frac{\sqrt{b}}{a} \cphi}- S^r_{213} \left(Z + \sqrt{b} \right) {\mathrm e}^{-2 \frac{\sqrt{b}}{a} \cphi} \right),\nonumber \\
	s_2^\cphi(r, \cphi, Z) & = & \frac{1}{8 a r^4} \left[-8 a r^3 \left( S^r_{212} \left(Z - \sqrt{b}\right) {\mathrm e}^{2 \frac{\sqrt{b}}{a} \cphi}+S^r_{213} \left(Z + \sqrt{b}\right) {\mathrm e}^{-2 \frac{\sqrt{b}}{a} \cphi} \right) + 4 a S^\cphi_{211} r^6 \right. + \nonumber \\ & & \left. + 8 a S^\cphi_{212} r^4 + \left(3 a^2 S^r_{120} - 4 S^r_{120} b - 8 S^r_{110} \left(Z^2 - b\right)\right) r^2 - 4 S^r_{120} \left(Z^2 - b\right)^2\right], \nonumber \\
	s_2^Z(r, \cphi, Z) & = & - \frac{2\sqrt{b} r}{a} \left(S^r_{212} {\mathrm e}^{2 \frac{\sqrt{b}}{a} \cphi} - S^r_{213} {\mathrm e}^{-2 \frac{\sqrt{b}}{a} \cphi}\right) , \nonumber 
\end{eqnarray}
where $S^r_{110}$, $S^r_{120}$, $S^r_{212}$, $S^r_{213}$, $S^\cphi_{10}$, $S^\cphi_{211}$ and $S^Z_{10}$
are integration constants (their designation corresponds to the function $s_j^{r,\cphi,Z}(r,\cphi,Z)$ in which they first arose in the process of solution).
For negative values of $b$ the solution~\eqref{oblateprolate_leading_order_solved}
shall be interpreted in terms of real trigonometric functions using the Euler formula ${\mathrm e}^{{\mathrm i} \kappa}=\cos(\kappa)+{\mathrm i} \sin(\kappa)$ and suitable complex choices of the integration constants $S^r_{212}$ and $S^r_{213}$ so that the resulting expressions in~\eqref{oblateprolate_leading_order_solved} are real;
for the special values of the parameters $b = -\frac{35 a^2}{12}$ and $b = -\frac{5 a^2}{4}$ certain terms in the compatibility conditions for $S_2^\cphi(r, Z)$ and $S_1^r(r, Z)$ vanish and the resulting solutions are somewhat different from~\eqref{oblateprolate_leading_order_solved}; however they lead in the end to the same conclusion as the generic case.

Clearly, if we are interested in solutions well defined in the cylindrical coordinates, for $b>0$ we should set $S^{r}_{212}=S^r_{213}=0$ to ensure the required periodicity in the angular variable $\cphi$; and restrict $2 \frac{\sqrt{-b}}{a}$ to integer values or set $S^{r}_{212}=S^r_{213}=0$ for $b<0$. However, even if we continue without this assumption, substituting~\eqref{oblateprolate_leading_order_solved} or its analogues into the lower order conditions and proceeding similarly as in the Appendix, we find that the only system with the integrals of the form~\eqref{oblateprolate} and nonvanishing magnetic field has the magnetic field and the electrostatic potential of the form
\begin{equation}\label{oblateprolate_full_solution}
	B_r(r, \cphi, Z) = 0, \quad B_\cphi(r, \cphi, Z) = 0, \quad B_Z(r, \cphi, Z) = b_\phi r, \quad W(r, \cphi, Z) = \frac{w_0}{r^2} - \frac{b_\cphi^2}{8} r^2,
\end{equation}
or, equivalently, in the Cartesian coordinates
\begin{equation}
	B_1(x,y,z)=0, \; B_2(x,y,z)=0, \; B_3(x,y,z)={b_\phi}, \; W(x,y,z) = \frac{w_0}{x^2+y^2} -\frac{b^2_\phi}{8}(x^2+y^2).
\end{equation}

This system is already known, see e.g. Class III in \cite{Marchesiello2018Sph}.
It possesses the commuting first order integrals of cylindrical type 
\begin{equation}
	p_Z^A, \quad p_\cphi^A - \frac{ b_\cphi}{2} r^2, 
\end{equation}
and another spherical type integral which can be combined together to construct the commuting second order integrals of the form~\eqref{oblateprolate}. Thus, no integrable system with integrals of the form~\eqref{oblateprolate} exists which would not belong also to another, standard class, namely cylindrical and spherical.\medskip

\subsection{The generalized circular parabolic cases}

We performed a similar analysis, and with the same negative result, also for the generalized circular parabolic case, i.e. integrals of the form~\eqref{circparabolic}. 

In this case it is again convenient to work in the cylindrical coordinates~\eqref{cyl coords}. Due to the assumed structure of the integrals~\eqref{circparabolic} the leading order determining equations coming from~\eqref{commutH} for the first integral are identical to the equations~\eqref{oblateprolateHX2}, after renaming $s_2^{r,\cphi,Z}(r,\cphi,Z)$ to $s_1^{r,\cphi,Z}(r,\cphi,Z)$. Thus also their solution is expressed as in~\eqref{Lz2_int_sol}, up to appropriate relabelling of the functions.

The remaining leading order determining equations read
\begin{eqnarray}\label{circparabolicHX2}
	\nonumber & & r B_\cphi(r, \cphi, Z) - \partial_r s_2^r(r, \cphi, Z)=0, \\ \nonumber & &
	r B_\cphi(r, \cphi, Z) - a B_r(r, \cphi, Z) + \partial_Z s_2^Z(r, \cphi, Z)=0,\\ \nonumber & &
	a r B_r(r, \cphi, Z) + r \partial_\cphi s_2^\cphi(r, \cphi, Z) + s_2^r(r, \cphi, Z)=0,\\ & & 
	a B_Z(r, \cphi, Z)- 2 Z B_\cphi(r, \cphi, Z) + \partial_r s_2^Z(r, \cphi, Z) + \partial_Z s_2^r(r, \cphi, Z)=0,\\ & & \nonumber 
	2 Z B_r(r, \cphi, Z)- r B_Z(r, \cphi, Z) + r^2 \partial_Z s_2^\cphi(r, \cphi, Z) + \partial_\cphi s_2^Z(r, \cphi, Z)=0,\\ & &\nonumber 
	r B_r(r, \cphi, Z)- a r^2 B_\cphi(r, \cphi, Z) + r^2 \partial_r s_2^\cphi(r, \cphi, Z) + \partial_\cphi s_2^r(r, \cphi, Z)=0
\end{eqnarray}
from $\{H,X_2 \}=0$, and
\begin{eqnarray}\label{circparabolicX1X2}
	\nonumber
	& & 
	r \partial_Z s_1^r(r, \cphi, Z) + s_1^Z(r, \cphi, Z)=0, 
	\\ \nonumber & & 
	a r^2 \partial_Z s_1^\cphi(r, \cphi, Z) + 2 r s_2^r(r, \cphi, Z) + s_1^Z(r, \cphi, Z)=0,
	\\ & & 
	- 2 a r B_r(r, \cphi, Z)+a r \partial_Z s_1^\cphi(r, \cphi, Z) + \partial_Z s_1^r(r, \cphi, Z)=0,
	\\ \nonumber & & 2 r^3 \partial_Z s_2^\cphi(r, \cphi, Z) + r^2 \partial_r s_1^\cphi(r, \cphi, Z) + 2 r Z \partial_Z s_1^\cphi(r, \cphi, Z) - a s_1^r(r, \cphi, Z)=0,
	\\ \nonumber & & 
	-2 a r^2 B_\cphi(r, \cphi, Z) + 2 r^2 \partial_r s_2^\cphi(r, \cphi, Z) + r \partial_Z s_1^\cphi(r, \cphi, Z) + a \partial_Z s_1^r(r, \cphi, Z)=0,
	\\ \nonumber & & 
	2 a r^2 B_Z(r, \cphi, Z) + a r^2 \partial_r s_1^\cphi(r, \cphi, Z) - 2 Z \partial_Z s_1^r(r, \cphi, Z) + s_1^r(r, \cphi, Z)=0
\end{eqnarray}
from $\{X_1,X_2 \}=0$ (where again the previous equations were used for simplification.) Equations~\eqref{circparabolicX1X2}  can be easily solved, specifying the dependence of $s_1^{r,\cphi,Z}$ and $s_2^{r,\cphi,Z}$ (and thus also of $B_Z(r,\cphi,Z)$ and $B_r(r,\cphi,Z)$) on the angular variable $\cphi$ in terms of constants of integration. The remaining conditions~\eqref{circparabolicHX2} determine the magnetic field component $B_\cphi(r,\cphi,Z)$ in terms of $s_2^{r}(r,\cphi,Z)$ and further constrain the dependence of $s_1^{r,\cphi,Z}$ and $s_2^{r,\cphi,Z}$ on the coordinates $r,Z$. For the generic values of $a,b$ we arrive at the solution
\begin{eqnarray}\label{circparabolicSBsol}
	B_r(r, \cphi, Z) & = & \frac{S^{\cphi}_{101}}{2 r^2}-\frac{(1+a^2) S^{Z}_{111}}{2 a^2 r} {\mathrm{e}}^{-\frac{\cphi}{a}} , \qquad B_\cphi(r, \cphi, Z) = \frac{a S^{\cphi}_{101}}{2 r^3}, \nonumber \\ 
	B_Z(r, \cphi, Z) & = & \frac{S^{\cphi}_{101} Z}{r^3}-\frac{2 S^Z_{201} r}{a} - \frac{1+a^2}{2 a^2 r^2 } \left( S^r_{101} {\mathrm{e}}^{\frac{\cphi}{a}}+S^{Z}_{111} Z {\mathrm{e}}^{-\frac{\cphi}{a}} \right), 
	\nonumber \\ 
	s_1^r(r, \cphi, Z) & = & \frac{1}{a} \left( S^r_{101} {\mathrm{e}}^{\frac{\cphi}{a}} -S^{Z}_{111} Z {\mathrm{e}}^{-\frac{\cphi}{a}} \right),\qquad s_1^Z(r, \cphi, Z) = \frac{S^{Z}_{111} r}{a} {\mathrm{e}}^{-\frac{\cphi}{a}}, \nonumber \\ 
	s_1^\cphi(r, \cphi, Z) & = & \frac{2 S^Z_{201} r^2}{a} + \frac{S^{\cphi}_{101} Z}{r^2} + S^\cphi_{100} -\frac{1}{r} \left(S^r_{101} {\mathrm{e}}^{\frac{\cphi}{a}}+S^{Z}_{111} Z {\mathrm{e}}^{-\frac{\cphi}{a}} \right), \nonumber \\ 
	s_2^r(r, \cphi, Z) & = & - \frac{a S^{\cphi}_{101}}{2 r}+\frac{(a^2 - 1) S^{Z}_{111}}{2 a} {\mathrm{e}}^{-\frac{\cphi}{a}}, 
	\\ 
	s_2^\cphi(r, \cphi, Z) & = & -\frac{2 S^Z_{201} Z}{a} - \frac{(a^2 - 1) S^{\cphi}_{101}}{4 r^2} + S^{\cphi}_{200} - \frac{S^{Z}_{111}}{r} {\mathrm{e}}^{-\frac{\cphi}{a}} , \nonumber \\ 
	s_2^Z(r, \cphi, Z) & = & S^Z_{201} r^2 + S^Z_{200} - \frac{1+a^2}{2 a r}\left(S^r_{101}{\mathrm{e}}^{\frac{\cphi}{a}}+ S^{Z}_{111} Z {\mathrm{e}}^{-\frac{\cphi}{a}} \right), \nonumber 
\end{eqnarray}
where $S^r_{101}$, $S^{\cphi}_{101}$, $S^\cphi_{100}$, $S^{\cphi}_{200}$, $ S^{Z}_{111}$, $S^Z_{201}$ and $S^Z_{200}$ are integration constants (as before, their designation corresponds to the function $s_j^{r,\cphi,Z}(r,\cphi,Z)$ in which they first arose in the process of solution).

For the particular values $a^2 =3$ and $a^2 = \frac{3}{5}$ the solution again somewhat differs but structurally is similar to~\eqref{circparabolicSBsol}.

As in the generalized oblate /prolate spheroidal cases, we can set $S^{Z}_{111}=S^r_{101}=0$ to ensure the required periodicity of the magnetic field in the angular variable, thus eliminating any dependence on $\cphi$ in~\eqref{circparabolicSBsol}. The solution of the lower order determining equations further constrains the parameters and we find again only the system~\eqref{oblateprolate_full_solution}, both for the generic and particular values of the parameter $a$.

\subsection{Case not directly related to any coordinate system}\label{sec:other}
In this section, we shall deal with the case~\eqref{nocoord}. The structure of the leading order terms in the integrals is not related to separation in any orthogonal coordinate system. However, since only $l_3$ and $p_3$ are present in the leading order terms, it is obvious that 
\begin{itemize}
	\item the cylindrical coordinates are best suited for the investigation of this case, and
	\item any system with the commuting first order cylindrical type integrals $l_3+\ldots$ and $p_3+\ldots$ belongs to the case~\eqref{nocoord} up to polynomial combination of its integrals. Since such systems were already considered before, cf.~\cite{Fournier2019,KMS2022}, we shall exclude them from our consideration at any step.
\end{itemize}
The leading order terms in the determining equations~\eqref{commutH} and~\eqref{commutX1X2} imply
\begin{eqnarray}\label{nocoordHX1}
	&& \partial_Z s_1^Z(r, \cphi, Z)=0, \qquad \partial_r s_1^r(r, \cphi, Z)=0,
	\nonumber \\ 
	&& 2 \left( a - r^2 \right) B_r(r, \cphi, Z)+\partial_\cphi s_1^Z(r, \cphi, Z) + r^2 \partial_Z s_1^\cphi(r, \cphi, Z) =0, 
	\nonumber \\ && - 2 a B_\cphi(r, \cphi, Z)+\partial_r s_1^Z(r, \cphi, Z) + \partial_Z s_1^r(r, \cphi, Z)=0, 
	\nonumber \\ 
	&& r \partial_\cphi s_1^\cphi(r, \cphi, Z) + s_1^r(r, \cphi, Z)=0, 
	\nonumber \\ 
	&& 2 r^2 B_Z(r, \cphi, Z) +\partial_\cphi s_1^r(r, \cphi, Z) + r^2 \partial_r s_1^\cphi(r, \cphi, Z) =0, 
\end{eqnarray}
from $\{H,X_1\}=0$;
\begin{eqnarray}\label{nocoordHX2}
	&& B_r(r, \cphi, Z)-\partial_Z s_2^Z(r, \cphi, Z) =0, \qquad \partial_r s_2^r(r, \cphi, Z)=0,
	\nonumber \\ 
	&& 2 b B_r(r, \cphi, Z)+ \partial_\cphi s_2^Z(r, \cphi, Z) + r^2 \partial_Z s_2^\cphi(r, \cphi, Z) =0,
	\nonumber \\ 
	&& B_Z(r, \cphi, Z) - 2 b B_\cphi(r, \cphi, Z) +\partial_r s_2^Z(r, \cphi, Z) + \partial_Z s_2^r(r, \cphi, Z)=0, 
	\nonumber \\ 
	&& r B_r(r, \cphi, Z) +r \partial_\cphi s_2^\cphi(r, \cphi, Z) + s_2^r(r, \cphi, Z)=0, 
	\nonumber \\ 
	&& - r^2 B_\cphi(r, \cphi, Z)+\partial_\cphi s_2^r(r, \cphi, Z) + r^2 \partial_r s_2^\cphi(r, \cphi, Z)=0
\end{eqnarray}
from $\{ H,X_2\}=0$; and finally
\begin{eqnarray}\label{nocoordX1X2}
	& & r \partial_Z s_1^\cphi(r, \cphi, Z) + 2 s_2^r(r, \cphi, Z)=0, 
	\nonumber \\ 
	& & 2r^3 \partial_Z s_2^\cphi(r, \cphi, Z) + 2 b r \partial_Z s_1^\cphi(r, \cphi, Z) - 2 a r \partial_Z s_2^\cphi(r, \cphi, Z) - s_1^r(r, \cphi, Z)=0, 
	\nonumber \\ 
	& & 2(r^2 - a) B_r(r, \cphi, Z) - r^2 \partial_Z s_1^\cphi(r, \cphi, Z) =0, 
	\nonumber \\ 
	& & - 2 r^2 B_\cphi(r, \cphi, Z)+\partial_Z s_1^r(r, \cphi, Z) + 2 r^2 \partial_r s_2^\cphi(r, \cphi, Z) =0, 
	\nonumber \\ 
	& & 2 r^2 B_Z(r, \cphi, Z) -2 b \partial_Z s_1^r(r, \cphi, Z) + 2 a \partial_Z s_2^r(r, \cphi, Z) + r^2 \partial_r s_1^\cphi(r, \cphi, Z) =0
\end{eqnarray}
from $\{X_1,X_2\}=0$ (simplified, as before, using~\eqref{nocoordHX1} and~\eqref{nocoordHX2}).

The solution of these equations proceeds rather similarly to the previous cases, solving first~\eqref{nocoordHX1} and then considering equations~\eqref{nocoordX1X2} and~\eqref{nocoordHX2}. In the process several splittings arise, leading to particular solutions when $a=0$, $b\neq 0$ and $a=b=0$. The generic solution, which holds if $a\neq0$, reads
\begin{eqnarray}\label{nocoordssol}
	B_r(r, \cphi, Z) & = & 0, \qquad B_\cphi(r, \cphi, Z) = \frac{\partial_r S^Z_1(r)}{2 a}, 
	\qquad 
	B_Z(r, \cphi, Z) = -\frac{ \partial_r S^\cphi_{11}(r)}{2}- r S^\cphi_{123}-\frac{ S^r_{133}}{2 r^2}, \nonumber \\ 
	s_1^r(r, \cphi, Z) & = & 0, \qquad
	s_1^\cphi(r, \cphi, Z) = S^\cphi_{11}(r) + \left( r^2 - a \right) S^\cphi_{123} - \frac{S^r_{133}}{r}, \qquad s_1^Z(r, \cphi, Z) = S^Z_1(r), 
	\nonumber \\
	s_2^r(r, \cphi, Z) & = & 0,\qquad s_2^\cphi(r, \cphi, Z) = S^\cphi_{211} + \frac{S^Z_1(r)}{2 a}, 
	\\
	s_2^Z(r, \cphi, Z) & = & \frac{b}{a} S^Z_1(r) + S^Z_{231} +\frac{1}{2}\left( S^\cphi_{11}(r) + r^2 S^\cphi_{123} - \frac{S^r_{133}}{r} \right), \nonumber
\end{eqnarray}
i.e. depends not only on the integration constants $S^r_{133}$, $S^\cphi_{211}$, $S^\cphi_{123}$ and $S^Z_{231}$ but involves also two arbitrary functions of a single variable, $S^\cphi_{11}(r)$ and $S^Z_1(r)$. However, solving the lower order determining equations we arrive at the system with the magnetic field as in~\eqref{nocoordssol} and
\begin{equation}
	W(r, \cphi, Z) = W(r). 
\end{equation}
Thus the system's magnetic field and electrostatic potential depend only on the radial coordinate $r$ and the corresponding rotational and translational symmetry imply that it possesses two commuting cylindrical type first order integrals
\begin{equation}
	p_Z^A + \frac{S_1^Z(r)}{2 a}, \qquad p_\cphi^A + \frac{S_{11}^\cphi(r)}{2}+ \frac{S^\cphi_{123}}{2} r^2 -\frac{S^r_{133}}{r}. 
\end{equation}
This means that its integrals of the form~\eqref{nocoord} are reducible to the standard cylindrical ones and the system is of no interest here. The same result is obtained in a similar way also for the particular cases $a=0$, $b\neq 0$ and $a=b=0$.

However, let us notice that systems with commuting pairs of integrals not directly related to any coordinates can exist \cite{HS2022}, albeit with structure of the leading order terms different from~\eqref{nocoord}.

\section{Conclusions and perspectives}\label{sec:concl}
This work constitutes a step forward in the classification problem of quadratically integrable systems with vector potentials. We studied the conditions for the existence of a pair of commuting quadratic integrals in four cases, namely generalized elliptic cylindrical, prolate / oblate spheroidal and circular parabolic integrals, together with a case not related to any coordinate system. In the first three cases the integrals are so called because their quadratic terms generalize the structure that would allow separation in the corresponding non-subgroup type coordinates, namely the elliptic cylindrical, prolate / oblate spheroidal and circular parabolic coordinates, respectively. 

We found new integrable systems only in the generalized elliptic cylindrical case, namely systems \eqref{BellipAnotZero}-\eqref{WellipAnotZero} and \eqref{mag_ell_a0}-\eqref{pot_ell_a0}. Even in the limit of vanishing magnetic field, these systems lose any direct relation to separability in elliptic cylindrical coordinates. In \eqref{BellipAnotZero}-\eqref{WellipAnotZero} for $\beta_1,\beta_2\to0$ and in \eqref{mag_ell_a0}-\eqref{pot_ell_a0} for $\alpha_1, \alpha_2, \beta_1, \beta_2\to0$ not only the magnetic field goes to zero, but also the potential $W$. Thus, both systems reduce to free motion and therefore are separable in every choice of orthogonal coordinates in the Euclidean space, not only the elliptic cylindrical ones. In all the cases studied so far, when generalized subgroup type integrals are considered, the limit for vanishing magnetic field leads to separation in a subgroup- type coordinate system, namely cylindrical or Cartesian coordinates \cite{KMS2022,Marchesiello2017}, even when the integrals do not generalize subgroup type ones \cite{HS2022}. Other possible subgroup type coordinates in the Euclidean space are the spherical coordinates, however in this case systems with a pair of generalized integrals cannot exist \cite{KMS2022}. On the other hand, four classes of systems with standard spherical type integrals exist \cite{Marchesiello2018Sph}. Their limit for vanishing magnetic field always leads to separation in spherical coordinates (or to multiseparability corresponding to another pair of quadratic integrals in the superintegrable exceptions); we never obtain free motion. As in other problems \cite{Marchesiello2015a}, there is some disparity between subgroup and non-subgroup type integrals, yet not well understood and worth to be investigated in the future.

Let us also notice that for $a=c=d=0$ the generalized elliptic cylindrical integrals would reduce to the standard ones. Systems admitting a pair of standard elliptic cylindrical integrals are (to our knowledge) not yet classified. Already in two dimension the problem was only partially addressed, cf.~\cite{Pucacco_2005}. Notice that if in the system introduced in section~\ref{sec:ellipAnotZero} we take the limit $c,d\to 0$ and next $a\to 0$, $a \beta_1 \to \tilde\beta_1 \in\mathbb{R}$ the integrals $X_1$ and $X_2$ have well defined limits, thus we arrive at a particular solution of the determining equations for the standard elliptic cylindrical integrals depending on 5 parameters. The system is determined by
\begin{eqnarray}
	\vec{B}(x,y,z) & = & \left(0, 0 , \tilde{\beta}_1-3 \beta_2 (x^2+y^2)\right),\label{limitstandellcyl}  \\
	W(x,y,z) &=& \beta_2 \left(  \frac{\tilde\beta_1}{4}  (x^2 + y^2)^2 +\frac{b \beta_2}{2} (x^4 - y^4) -  \frac{\beta_2}{4} (x^2 + y^2)^3 - b \tilde\beta_1  x^2   +  \omega_3 (x^2 + y^2) + \omega_1 x  +  \omega_2 y  \right).\nonumber 
\end{eqnarray}
As we already mentioned in section~\ref{sec:ellipAnotZero}, in a suitably chosen gauge this system has an integral $\tilde{X}_2=p_3$. For the magnetic field and potential as in~\eqref{limitstandellcyl} the degrees of freedom actually decouple in the $z$ and $xy$ directions, the system~\eqref{limitstandellcyl} can be constrained to the $xy$ plane where it becomes of the 2D elliptic type. Some systems of this type were constructed in~\cite{Pucacco_2005}; however by inspection we find that~\eqref{limitstandellcyl} has different structure of the magnetic field and potential and thus provides us with another solution of this still not fully classified 2D problem and an example of a system with standard pair of elliptic cylindrical integrals.

For the system \eqref{mag_ell_a0}-\eqref{pot_ell_a0} of section~\ref{sec:ellipAZero} we have $\delta=\frac{2\sqrt{b}}{c}$. We can remove the singularity for $c\to0$ by setting $b= c^2 b_1$, however then the integral $X_1$ would reduce to generalized cylindrical one, cf.~\cite{KMS2022}, and not to standard elliptic cylindrical integrals. 

Let us observe that attempting to directly solve the determining equations for $X_1$ and $X_2$ in the case $b\neq 0$, $a=c=d=0$, one arrives at a system of differential equations which is less overdetermined than~\eqref{eqs21-1}--\eqref{eqs2-6}, thus one has less compatibility conditions for them at his disposal and the equations appear to be much more difficult to solve in full generality. However, as the focus of this paper is on generalized pairs of integrals, consideration of this standard class goes beyond its scope and we defer it to future work.

For potential applications, we find the new integrable system~\eqref{mag_ell_a0}--\eqref{pot_ell_a0} with $b<0$ of most interest due to its periodicity in the $z$--coordinate, as the periodic magnetic fields appear in numerous applications, see e.g.~\cite{BALAL2020163895,HeinIld,LACROIX2009906,SINITSYN200225,HUSE_Optik}. Thus more detailed analysis of its properties, e.g. how generic are its bounded trajectories, cf. Fig~\ref{pic2}, is worth pursuing. 

So far, no maximally superintegrable system has been found admitting a pair of commuting generalized integrals which would not be reducible to functions of the standard ones. One minimally superintegrable system is known, namely a helical undulator placed in an infinite solenoid \cite{KMS2022}, possessing a pair of generalized cylindrical integrals and a pair of Cartesian integrals (not all independent with the Hamiltonian). Despite being minimally superintegrable, this systems does not separate in the configuration space. The investigation of the eventual superintegrability of the newly found integrable systems \eqref{BellipAnotZero}-\eqref{WellipAnotZero} and \eqref{mag_ell_a0}-\eqref{pot_ell_a0} is currently work in progress.

\appendix
\section{Solution of the determining equations for the generalized elliptic cylindrical case}\label{sec:appendix}
Let us show in detail the solution of the determining equations that lead to the integrable systems \eqref{BellipAnotZero}-\eqref{WellipAnotZero} and \eqref{mag_ell_a0}-\eqref{pot_ell_a0} in the following sections \ref{sec:anotzero-comp} and \ref{sec:azero-comp}, respectively.

\subsection{$a\neq0$}\label{sec:anotzero-comp}
If $a\neq0$, equation \eqref{eq:split-a} implies $\partial_z S_{22}(z)=0$, therefore $S_{22}(z)=S_{22}$, $S_{22}\in\mathbb R$. By substituting into \eqref{eqs2-1} we conclude that necessarily
\begin{equation}\label{S212sol} 
	S_{21}^2(z)= -\frac{1}{2} a \partial_z S_{21}^1(z).
\end{equation}
Equation \eqref{eqs2-2} then implies
$$\frac{1}{2} a^2 \partial_z^2 S_{21}^1(z)+2 S_{21}^1(z)=0,$$
solved by
\begin{equation}\label{S211sol}
	S_{21}^1(z)=\alpha_1 \sin \left(\frac{2 z}{a}\right)+\alpha_2\cos \left(\frac{2 z}{a}\right),\;\; \alpha_i\in\mathbb R.
\end{equation}
By substituting \eqref{S212sol} and \eqref{S211sol} into \eqref{eqs2-1}, \eqref{eqs2-2} and \eqref{eqs2-4}
we obtain 
\begin{eqnarray}
	&& c \left( \alpha_1 \cos \left(\frac{2 z}{a}\right)- \alpha_2 \sin \left(\frac{2 z}{a}\right)\right)=0, \label{eqs2last1} \\
	&& d\left(\alpha_1 \sin \left(\frac{2 z}{a}\right)+\alpha_2 \cos \left(\frac{2 z}{a}\right)\right)=0,\\
	&&c \left(\alpha_1 \sin \left(\frac{2 z}{a}\right)+\alpha_2\cos \left(\frac{2 z}{a}\right)\right)+d \left( \alpha_1\cos \left(\frac{2 z}{a}\right)-\alpha_2 \sin \left(\frac{2 z}{a}\right)\right)- a b S_{22}=0.\label{eqs2last3}
\end{eqnarray}
We are thus lead to a second level splitting, depending on the values of the constants $c$ and $d$.

\subsubsection{$ c^2+ d^2 \neq 0$}
If $c$ or $d$ is not zero, we see that \eqref{eqs2last1}-\eqref{eqs2last3} can be satisfied only if $\alpha_1=\alpha_2=S_{22}=0$ (both $a$ and $b$ cannot be zero). This implies that
\begin{eqnarray}
	&& s_2^1(x,y,z)=s_2^2(x,y,z)=0,\;\;s_2^3(x,y,z)=S_2^3(x,y), \label{solX22Sa}\\
	&& B_1(x,y,z)= -\frac{1}{2} \partial_y S_2^3(x,y),\;\;B_2(x,y,z)= \frac{1}{2} \partial_x S_2^3(x,y),\label{solB12Sa}
\end{eqnarray}
while the remaining second order equations \eqref{eqs2-5}-\eqref{eqs2-6}
simplify to 
$$\partial_z s_1^1(x,y,z)=\partial_z s_1^2(x,y,z)=0$$
and are therefore solved by
\begin{equation}\label{solX12Sa}
	s_1^1(x,y,z)= S_1^1(x,y),\;\; s_1^2(x,y,z)= S_1^2(x,y).
\end{equation}
At this point, both second order conditions for the integral $X_2$ and the commutativity condition \eqref{commutX1X2} are solved. 

From \eqref{eqs21-2} we get
\begin{equation}\label{solB3Sa}
	B_3(x,y,z)=\frac{(a x+d) \partial_y S_2^3(x,y)-2\partial_y S_1^2(x,y)}{4 x y}.
\end{equation}
The remaining second order equations for $X_1$ are quite complicated. Let us continue with the simpler first order conditions coming from \eqref{commutH} for the integral $X_2$, namely
\begin{eqnarray}
	S_2^3(x,y) \partial_x S_2^3(x,y)-2\partial_x m_2(x,y,z)=0, \label{eqX21-1}\\
	S_2^3(x,y) \partial_ yS_2^3(x,y)-2\partial_y m_2(x,y,z)=0, \label{eqX21-2}\\
     2 \partial_z W(x,y,z)- 	\partial_z m_2(x,y,z)=0, \label{eqX21-3}
\end{eqnarray}
where we substituted \eqref{solX22Sa}, \eqref{solB12Sa} and \eqref{solB3Sa}. 

Since $\partial_{xz} m_2=\partial_{z x} m_2$ and $\partial_{yz} m_2=\partial_{z y} m_2$, the above equations imply $\partial_{xz} W= \partial_{yz} W=0$ and therefore
\begin{equation}
	W(x,y,z)=W_1(x,y)+ W_2(z).
\end{equation}
Thus, \eqref{eqX21-1}--\eqref{eqX21-3} are solved by
\begin{equation}\label{m2sola}
	m_2(x,y,z)=\frac{1}{4} s_2^3(x,y)^2+2 W_2(z).
\end{equation}
After substituting \eqref{solX22Sa}, and \eqref{m2sola}, the first order equations for the involutivity condition \eqref{commutX1X2} simplify to
\begin{equation}
	-2 (c-a y) W_2'(z)=0,\;\;-2 (a x+d) W_2'(z)=0,\;\; \partial_z m_1(x,y,z)=0.
\end{equation}
Therefore, 
\begin{equation}\label{m1Sola}
	m_1(x,y,z)= M_1(x,y)
\end{equation}
and since $a\neq0$, necessarily $ W_2(z)=0$. 

By using all we know so far about the solutions for $s_2^j$ and $m_j$, we see that the zero order equations coming from the integral $X_2$ and from the involutivity relation \eqref{commutX1X2} are identically zero. In particular, $X_2$ takes the form
\begin{equation}\label{X2Sola}
	X_2= \frac12(p_3^A+ s_2^3(x,y))^2
\end{equation}
and if the gauge is chosen so that $A_3(x,y,z)=- s_2^3(x,y)$ it simplifies to $X_2=p_3$. This reflects the fact that the $z$ coordinate is cyclic and indeed
\begin{equation}\label{WSola}
	W(x,y,z)=W_1(x,y).
\end{equation}

However, we still have to solve the determining equations for the integral $X_1$, except \eqref{eqs21-2}. Let us start by the remaining second order conditions from \eqref{eqs21-1}-\eqref{eqs21-6}. 
We find it convenient to change the coordinate system to a shifted and scaled one, namely
\begin{equation}\label{XYshift}
	x=\frac{X-d}{a},\;\;y= \frac{c+Y}{a},
\end{equation}
so that \eqref{eqs21-1}-\eqref{eqs21-6} reduce to (where \eqref{eqs21-2} was already solved by \eqref{solB3Sa})
\begin{eqnarray}
	&&X \partial_Y S_{2}^3(X,Y)-Y \partial_X S_{2}^3(X,Y)=0, \label{eq2X1-1}\\
	&& 2(\partial_X S_{1}^1(X,Y)+\partial_Y S_{1}^2(X,Y))- X \partial_Y S_{2}^3(X,Y)+ Y\partial_X S_{2}^3(X,Y)=0,\label{eq2X1-2}\\
	&& -2 a^4 Y\partial_y S_{1}^2(X,Y)-4 (c+Y) (d-X) \left(a^2 \partial_y S_{1}^3(X,Y) +(c+Y) (d-X) \partial_x S_{2}^3(X,Y)\right)\nonumber\\
	&&+\left(a^4 X Y-4 (d-X)^3 (c+ Y)\right)\partial_Y S_{2}^3(X,Y)=0, \label{eq2X1-3} \\
	&& 4 (c+Y) (X-d) \left(a^2\partial_X S_{1}^3(X,Y)+ \partial_X S_{2}^3(X,Y)\left((c+Y)^2+a^2 b\right)\right) \nonumber\\
	&&-2 a^4 X \partial_Y S_{1}^2(X,Y)+\left(a^4 X^2-4 (c+Y)^2 (d-X)^2\right)\partial_Y S_{2}^3(X,Y)=0, \label{eq2X1-4}\\
	&&2\partial_Y S_{1}^2(X,Y) \left( a^2 b+(c+Y)^2- (d-X)^2\right)\nonumber \\
	&&+2 (c+Y) (X-d)\left( \partial_Y S_{1}^1(X,Y)
	+\partial_X S_{1}^2(X,Y)-\frac12 X\partial_X S_{2}^3(X,Y)\right) \nonumber \\
	&& +\left(-a^2 b X-c Y (d+X)-X (c+d-X) (c-d+X)-d Y^2\right)\partial_Y S_{2}^3(X,Y)=0 \label{eq2X1-5}.
\end{eqnarray}
The first equation above implies 
\begin{equation}\label{S23Solpolara}
	S_{2}^3(X,Y)= F(r),
\end{equation}
if polar coordinates
\begin{equation}\label{polar}
	X= r \cos (\cphi),\;\; Y=r \sin(\cphi),\;\;r\geq0,\;\; 0\leq\cphi<2\pi,
\end{equation}
are introduced. 
Though it seems more convenient to work in polar coordinates, the remaining equations do not simplify in polar coordinates. However, if we multiply the third equation above by $X$ and subtract from it the fourth equation multiplied by $Y$ and then transform the expression so obtained into polar coordinates we arrive at
\begin{eqnarray}
	&&r \partial_r F(r) \left(a^2 b \sin (2 \cphi)+c^2 \sin (2 \cphi)-2 c d \cos (2 \cphi)+2 c r \cos (\cphi)-d^2 \sin (2 \cphi)+2 d r \sin (\cphi)\right)\nonumber\\
	&&-2 a^2 \partial_{\cphi} S_1^3(r,\cphi)=0,
\end{eqnarray}
solved by
\begin{equation}\label{S13Solpolara}
	S_1^3(r,\cphi)=G(r)-\frac{r \partial_r F(r) \left(\cos (2 \cphi) \left(a^2 b+c^2-d^2\right)+4 d \cos (\cphi) (c \sin (\cphi)+r)-4 c r \sin (\cphi)\right)}{4 a^2}.
\end{equation}
Now let us continue in the $(X,Y)$ coordinates. We solve \eqref{eq2X1-1} for $\partial_X S_2^3(X,Y)$, namely $\partial_X S_2^3(X,Y)=\frac{X}{Y} \partial_Y S_2^3(X,Y)$ and substitute it into the remaining equations \eqref{eq2X1-2}--\eqref{eq2X1-5}, for simplicity let these be in the following $Q_j=0$, $j=1,\ldots 4$, respectively. 
We solve $Q_2=0$ for $\partial_Y S_1^2$ and together with $\partial_X S_2^3(X,Y)=\frac{X}{Y} \partial_Y S_2^3(X,Y)$ we substitute it into 
\begin{equation}\label{compEq2X1polar}
	\partial_Y(\partial_Y Q_1- \partial_X (((-d + X) (c + Y))^{-1}Q_4).
\end{equation} In this way we manage to arrive at an equation involving only $S_1^3$ and $S_2^3$. 
%
Once expressed in polar coordinates, this gives us an equation for the functions $F(r)$ and $G(r)$, cf. \eqref {S23Solpolara} and \eqref{S13Solpolara}.
We then expand the trigonometric expressions in \eqref{compEq2X1polar} to obtain a polynomial in $\sin(\cphi)$, $\sin(\cphi)\cos(\cphi)$, $\sin(\cphi)\cos^2(\cphi)$, $\cos^3(\cphi)$ and $\cos^4(\cphi)$ with coefficients depending on $r$. Equation \eqref{compEq2X1polar} is satisfied if all such coefficients are equal to zero. This corresponds to a system of equations that the functions $F$, $G$ have to satisfy. If we look at the coefficient of $\cos^4(\cphi)$, we get
\begin{equation}\label{eqF}
	c d \left(a^2 b+c^2-d^2\right) \left(15 F'(r)-r \left(15 F''(r)+r \left(r F^{(4)}(r)-6 F'''(r)\right)\right)\right)=0.
\end{equation}
If $c d \left(a^2 b+c^2-d^2\right)\neq0$, equation \eqref{eqF} implies
\begin{equation}\label{SolF}
	F(r)=\frac{1}{6} r^6 S^3_{23}+\frac{1}{4} r^4 S^3_{22}+\frac{1}{2} r^2 S^3_{21}+S^3_{20},\;\; S^3_{2j}\in\mathbb R.
\end{equation}
By substituting \eqref{SolF} into \eqref{compEq2X1polar} and looking at the coefficient of $\sin(\cphi)\cos(\cphi)$ we obtain the condition
\begin{eqnarray}
	&&-16 \left(a^2 \left(a^2 b+c^2-d^2\right) \left(3 G'(r)+r \left(r G^{(3)}(r)-3 G''(r)\right)\right)+2 r^5 \left(2 S^3_{23} \left(-7 r^2 \left(a^2 b-5 c^2+5 d^2\right.\right.\right.\right.)\nonumber\\
	&&+\left.\left.\left.\left(a^2 b+c^2\right)^2-d^4\right)-5S^3_{22} \left(a^2 b-3 c^2+3 d^2\right)\right)\right)=0, 
\end{eqnarray}
solved by
\begin{eqnarray}
	G(r) & = & \frac{S^{3}_{11}}{4} r^4 + \frac{S^{3}_{10}}{2} r^2 + S^3_{12} + 5 S^{3}_{22} \frac{a^2 b - 3 c^2 + 3 d^2}{24 a^2 (a^2 b + c^2 - d^2)} r^6 \nonumber \\ & & + S^{3}_{23} r^6 \left( 7 \frac{a^2 b - 5 c^2 + 5 d^2}{48 a^2 (a^2 b + c^2 - d^2)} r^2 -\frac{a^2 b + c^2 + d^2}{12 a^2} \right),\label{SolG}
\end{eqnarray}
where $S^3_{ij}\in\mathbb R$. By substituting also \eqref{SolG} into \eqref{compEq2X1polar}, we obtain that the remaining coefficients of trigonometric expressions vanish only if
\begin{equation}\label{SolGimplications}
	S^3_{21}=\frac{2}{3} a^2 S^3_{11},\; S^3_{22}= 0,\; S^3_{23}=0.
\end{equation}
Transforming back to the $(X,Y)$ coordinates, we therefore arrive at
\begin{eqnarray}
	S_2^3(X,Y)&=&\frac{1}{3} a^2 S^3_{11} \left(X^2+Y^2\right)+ S^3_{20}, \label{solS23}\\
	S_1^3(X,Y)&=&\frac{1}{12} S^3_{11} \left(2 Y^2 \left(a^2 b+c^2-d^2-4 d X+3 X^2\right)+X^2 \left(-2 a^2 b-2 c^2+2 d^2-8 d X+3 X^2\right)\right.\nonumber\\
	& & +\left. 8 c X Y (X-d)+8 c Y^3+3 Y^4\right)+\frac{1}{2} S^3_{10}\left(X^2+Y^2\right)+S^3_{12},\label{SolS13}
\end{eqnarray}
with $S^3_{1j}$, $S^3_{2 i}$ $\in\mathbb R$. Since $X_2$ is given by \eqref{X2Sola}, we can set $ S^3_{20}=S^3_{12}=0$ in the following without loss of generality (i.e. subtracting the first order integral $p_3^A+ S_2^3(X,Y)$ from $X_1$).
By substituting into \eqref{eqs21-1}-\eqref{eqs21-6}, we are finally left with equations for $S_1^1$ and $S_1^2$ that can be solved. Namely, we find
\begin{eqnarray}\label{SolS1XY}
	S_1^1(X,Y)&=&S_{11}^1-\frac{1}{6 a^2}\left(6 a^2 Y (2 b S^3_{10}+S_{12}^2)+S^3_{11} \left(6 c Y^2 \left(3 a^2 b+3 c^2+d^2-6 d X+3 X^2\right)\right.\right.\nonumber\\
	& & - c X (2 d-X) \left(2 \left(a^2 b+c^2+d^2\right)- 6 d X+3 X^2\right)+Y^3 \left(a^4+6 a^2 b+2 \left(13 c^2+d^2-6 d X+3 X^2\right)\right)\nonumber\\
	& & +Y \left(-4 d X \left(a^2 b+7 c^2+d^2\right)+4 \left(a^2 b+c^2\right)^2+X^2 \left(a^4+2 a^2 b+14 \left(c^2+d^2\right)\right)\right.\nonumber\\
	& & - \left.\left. 4 d^4-12 d X^3+3 X^4\right)+15 c Y^4+3 Y^5\right)+6 S^3_{10} Y \left((c+Y) (2 c+Y)-2 d^2-2 d X+X^2\right)\nonumber\\
	& & +\left. 6 c S^3_{10} X (X-2 d)\right)
\end{eqnarray}
and
\begin{eqnarray}\label{SolS2XY}
	S_1^2(X,Y)&=&\frac{S_{10}^3 \left(-d \left(Y (2 c+Y)+3 X^2\right)+X Y (2 c+Y)+X^3\right)}{a^2}+S_{12}^2 X+S_{11}^2\nonumber\\
	& & +\frac{1}{6 a^2}\left(X^2 \left(a^4 X+a^2 b (6 d-2 X)+2 c^2 (X-3 d)-18 d^3+26 d^2 X-15 d X^2+3 X^3\right)\right.\nonumber\\
	& & +4 c Y (X-d) \left(a^2 b+c^2+d^2-6 d X+3 X^2\right)\nonumber\\
	& +&\left. Y^2 \left(a^4 X+2 a^2 b (X-d)-2 (d-X) \left(7 c^2+d^2-6 d X+3 X^2\right)\right)+12 c Y^3 (X-d)\right.\nonumber\\
	& & + \left. 3 Y^4 (X-d)\right)S_{11}^3,
\end{eqnarray}
where $S_{1j}^i\in\mathbb R$.

All second order equations for the integral $X_1$ are now solved. After substituting the solution found for $W$ and $m_1$, cf. \eqref{WSola}, \eqref{m1Sola}, we have
\begin{eqnarray}
	&&\frac{2 \left(a^2 b+(c+Y)^2\right)\partial_X W_1(X,Y)}{a^2}+\frac{a^2 S_1^2(X,Y) \left(2 \partial_X S_1^1(X,Y)+X \partial_Y S_2^3(X,Y)\right)}{4 (c+Y) (d-X)}\nonumber\\
	&&+\frac{2 (c+Y) (d-X) \partial_Y W_1 (X,Y)}{a^2}+\frac{X \partial_Y S_2^3(X,Y) S_1^3(X,Y)}{2 Y}-\partial_x M_1(X,Y)=0,\label{eqX11-1}\\
	&&\frac{1}{4} \left(\frac{a^2 S_1^1(X,Y) \left(2\partial_X S_1^1 (X,Y)+X \partial_Y S_2^3(X,Y)\right)}{(c+Y) (d-X)} -2 \partial_Y S_2^3(X,Y) S_1^3(X,Y)\right.\nonumber\\
	&&\left.-\frac{8 (d-X) \left((c+Y) \partial_X W_1(X,Y)+(d-X) \partial_Y W_1(X,Y)\right)}{a^2}\right)+\partial_Y M_1(X,Y)=0,\label{eqX11-2}\\
	&&\frac{1}{2} \left(\frac{X \partial_YS_2^3(X,Y)S_1^1(X,Y)}{Y}+\partial_Y S_2^3(X,Y) S_1^2(X,Y)\right.\nonumber\\
	&&\left.-2 X\partial_Y W_1(X,Y)+2 Y \partial_X W_1(X,Y)\right)=0\label{eqX11-3}
\end{eqnarray}
and 
\begin{equation}\label{eqX10}
	S_1^1(X,Y)\partial_X W_1(X,Y)+S_1^2(X,Y) \partial_Y W_1(X,Y)=0
\end{equation}
for the first and zero order equations, respectively. For consistency with equations \eqref{eq2X1-1}--\eqref{eq2X1-5}, we express the above equations in the $(X,Y)$ coordinates, though it is easier to solve them in the polar coordinates \eqref{polar}. The functions $S_i^j$ are known from \eqref{SolS1XY}-\eqref{SolS2XY}, therefore the only unknowns are $M_1$ and the 
potential $W_1$. It is best to start with equation \eqref{eqX11-3}, that expressed in the polar coordinates \eqref{polar} would involve only $\partial_{\varphi} W$ and known functions, thus it can be easily integrated. In this way we find $W$ up to an arbitrary function of $r$ that is determined by \eqref{eqX11-3}.
Finally, the function $m_1$ can be found by solving the remaining equations \eqref{eqX11-1}-\eqref{eqX11-2}. By omitting constant terms in the potential $W$ and in $m_1$, we arrive at the solution given in section \ref{sec:ellipAnotZero}, where the constants have been renamed according to
\begin{equation*}
	\beta_1=\frac{1}{3} S_{11}^3 \left(2 \left(c^2+d^2\right)-a^2 b\right)-S_{10}^3,\;\;\beta_2=\frac{1}{3} a^3S_{11}^3,
\end{equation*}
\begin{eqnarray*}
	\omega_1&=&\frac{6 \beta_2 d^3 \left(c^2+d^2\right)}{a^5}-\frac{4 b \beta_2 d^3}{a^3}-\frac{2 \beta_1 d^3}{a^2}-\frac{\beta_2 d \left(c^2+d^2\right)}{2 a}-d S_{12}^2-S_{11}^2,\\
	\omega_2&=&-\frac{6\beta_2 c d^2 \left(c^2+d^2\right)}{a^5}+2 b c \left(\frac{\beta_2 \left(c^2+3 d^2\right)}{a^3}-\beta_1\right)+\frac{2 \beta_1 c d^2}{a^2}+\frac{\beta_2 c \left(c^2+d^2\right)}{2 a}+c S_{12}^2+S_{11}^1,\\
	\omega_3&=&\frac{a^5 (4 b \beta_1-2 S_{12}^2)-a^4 \beta_2\left(3 c^2+d^2\right)-2 a^3 \beta_1 \left(c^2+3 d^2\right)-12 a^2 b \beta_2 d^2+3 \beta_2\left(c^2+d^2\right) \left(c^2+5 d^2\right)}{4 a^4}.
\end{eqnarray*}


If $c d \left(a^2 b+c^2-d^2\right)=0$, equation \eqref{eqF} vanishes identically. We look at the coefficient of $\sin(\cphi) \cos(\cphi)^3$ in~\eqref{compEq2X1polar} which implies
\begin{equation}\label{eqFbis}
	\left((a^2 b + c^2 - d^2)^2 - 4 c^2 d^2\right) \left(15 F'(r)-r \left(15 F''(r)+r \left(r F^{(4)}(r)-6 F^{(3)}(r)\right)\right)\right)=0.
\end{equation}
Thus unless we have also $(a^2 b + c^2 - d^2)^2 - 4 c^2 d^2=0$ the equation~\eqref{SolF} determining $F(r)$ still holds. Similarly, the remaining conditions coming from~\eqref{compEq2X1polar} after substituting the solution for $F(r)$, cf.~\eqref{SolF}, imply that if $(a^2 b + c^2 - d^2)^2 - 4 c^2 d^2\neq 0$, we also must have 
\begin{equation}
	G(r) = \frac{ S^{3}_{11} }{4} r^4 + \frac{ S^{3}_{10} }{2} r^2 + S^3_{12}
\end{equation}
and
\begin{equation}
	S^3_{21}=\frac{2}{3} a^2 S^3_{11},\; S^3_{22}= 0,\; S^3_{23}=0;
\end{equation}
i.e. the computation in this case proceeds exactly as above.

The equations characterizing the case when both \eqref{eqF} and \eqref{eqFbis} identically vanish, namely
\begin{equation}
	c d \left(a^2 b+c^2-d^2\right)=0,\qquad (a^2 b + c^2 - d^2)^2 - 4 c^2 d^2=0,
\end{equation}
are equivalent to the equations
\begin{equation}
	c d=0, \qquad a^2 b+c^2-d^2=0,
\end{equation}
which have two solutions, $c=0$, $b=\frac{c^2}{a^2}$ or $d=0$, $b=-\frac{c^2}{a^2}$. However, they are related by a rotation of the coordinate frame around the $z$--axis by the angle $\frac{\pi}{2}$ accompanied by an appropriate linear combination of the integrals $H$, $X_1$ and $X_2$. Thus we proceed considering only the case $d=0$, $b=-\frac{c^2}{a^2}$.

In this case, the equation~\eqref{compEq2X1polar} significantly simplifies and implies only two independent conditions on the functions $F(r)$ and $G(r)$, namely
\begin{eqnarray}
	r^3 F^{(4)}(r) + 5 r^2 F'''(r) + 2 r F''(r) - 2 F'(r) & = & 0, \nonumber \\ 
	a^2 \left( r^2 G'''(r)+ r G''(r) - G'(r) \right) -r^4 F'''(r) - 6 r^3 F''(r) - 6 r^2 F'(r) & = & 0.
\end{eqnarray}
Their general solution reads
\begin{eqnarray}
	F(r) & = & S^3_{21} \ln(r) + S^3_{22} r^2 + \frac{S^3_{23}}{r}+ S^{3}_{20}, \\ \nonumber 
	G(r)& = & \frac{S^3_{10}}{2} r^2 + S^3_{11} \ln(r) + S^3_{12}+\frac{3 S^3_{22}}{4 a^2}r^4 + \frac{S^3_{21}}{2 a^2} r^2 \left(\ln(r) - 1 \right).
\end{eqnarray}
Thus we have determined the functions $S^3_2(r,\cphi)$ and $S^3_1(r,\cphi)$ in terms of seven integration constants. Continuing our analysis of the equations~\eqref{eq2X1-1}--\eqref{eq2X1-5} in polar coordinates~\eqref{polar}, we determine also $S^1_1(r,\cphi)$ and $S^2_1(r,\cphi)$, introducing three more integration constants and fully solving the equations~\eqref{eq2X1-1}--\eqref{eq2X1-5}.

The consistency of the lower order conditions requires
\begin{equation}
	S^3_{11}=S^3_{21}=0
\end{equation}
and leads to the splitting into two cases:
either 
\begin{equation}
	S^3_{23}=0,
\end{equation}
which after solution of all the lower order equations implies the solution of the form introduced in section \ref{sec:ellipAnotZero} for the particular values $d=0$, $b=-\frac{c^2}{a^2}$;
or
\begin{equation}\label{dzerosplitB}
	S^3_{22}=0.
\end{equation}
The system corresponding to the case~\eqref{dzerosplitB} is characterized by
\begin{eqnarray}
	B_x(x, y, z) & = & \frac{\beta \left( y - \frac{c}{a} \right)}{\sqrt{x^2 + \left( y - \frac{c}{a} \right)^2 }^3}, \qquad 
	B_y(x, y, z) = -\frac{\beta x}{\sqrt{ x^2 + \left( y - \frac{c}{a} \right)^2 }^{3}}, \\ \nonumber 
	B_z(x, y, z) & = & 0, \qquad 
	W(x, y, z) = \frac{\omega}{\sqrt{x^2 + \left( y - \frac{c}{a} \right)^2}} - \frac{\beta^2}{4 (x^2 + \left( y - \frac{c}{a} \right)^2) };
\end{eqnarray}
however, it possesses two commuting first order integrals
\begin{equation}
	X_3=l_Z^A, \quad \tilde{X}_2=p_Z^A+\frac{\beta}{2 \sqrt{x^2 + \left(y - \frac{c}{a} \right)^2}}
\end{equation}
of standard cylindrical type together with a pair of standard pair of parabolic cylindrical integrals, and therefore it is contained in~\cite{Kubu_2021}, only expressed in a shifted coordinate system. Thus, the generalized elliptical integral $X_1$ can be expressed as a function of the standard integrals.

\subsubsection{$c=d=0$}\label{sec:cdzero-comp}
If $c=d=0$ then the structure of the leading order terms hints that the equations are easiest to investigate in the cylindrical coordinates~\eqref{cyl coords}. Expressing the conditions~\eqref{eqs21-1}--\eqref{eqB2} and~\eqref{eqs2-1}--\eqref{eqs2-6} in these coordinates, it is rather straightforward if somewhat tedious, to solve them. The easiest approach is to solve first the conditions coming from the leading order terms of $\{ H,X_2 \}=0$, next the involutivity condition $\{ X_1,X_2 \}=0$ and finally use the leading order terms $\{ H,X_1 \}=0$, under the assumption $a\neq0$ and $b\neq 0$. We arrive at a general solution of the form
\begin{eqnarray}\label{an0cd0leadingsoln}
	B_r\left(r, \cphi, Z\right) & = & \frac{r}{a} \left( S^r_{212} \sin\left(\frac{2 Z}{a} - \cphi\right) -S^r_{211} \cos\left(\frac{2 Z}{a} - \cphi\right) \right), 
	\\ \nonumber 
	B_\cphi\left(r, \cphi, Z\right) & = & 3 S^Z_{21} r^5 + 2 S^Z_{22} r^3 + S^Z_{23} r - \frac{S^r_{211}}{a} \sin\left(\frac{2 Z}{a} - \cphi\right) - \frac{S^r_{212}}{a} \cos\left(\frac{2 Z}{a} - \cphi\right), 
	\\ \nonumber 
	B_Z\left(r, \cphi, Z\right) & = & \frac{1}{a} \left[ -S^Z_{101} r^3 - \frac{S^Z_{102}}{a} r + S^Z_{21} r^5 \left(6 b \cos\left(2 \cphi\right) - 7 r^2\right) + \right. \\ & & \nonumber + \left. S^Z_{22} r^3 \left(2 b \cos\left(2 \cphi\right) - 5 r^2 - 2 b\right) - S^Z_{23} b r\right] , 
	\\ \nonumber 
	s_1^r\left(r, \cphi, Z\right) & = & -\frac{b S^Z_{101}}{3 a} r^3 \sin\left(2 \cphi\right) - \frac{b S^Z_{102}}{a} r \sin\left(2 \cphi\right) + S^r_{110} \cos\left(\cphi\right) + S^r_{120} \sin\left(\cphi\right) + \\ \nonumber & & + \frac{b S^Z_{21}}{5 a} r^5 \left(-5 r^2 \sin\left(2 \cphi\right) + 3 b \sin\left(4 \cphi\right)\right) + \frac{b S^Z_{22}}{3a} r^3 \left(2 b \cos\left(2 \cphi\right) - 3 r^2 - 2 b\right) \sin\left(2 \cphi\right) - \\ \nonumber & & - \frac{b^2 S^Z_{23}}{a} r \sin\left(2 \cphi\right) - \frac{S^r_{211}}{4} \left(\left(a^2 + 2 b\right) \cos\left(\frac{2 Z}{a}-\cphi\right) + 2 b \cos\left(\frac{2 Z}{a}+\cphi \right)\right) + \\ \nonumber & & + \frac{S^r_{212}}{4} \left(\left(a^2 + 2 b\right) \sin\left(\frac{2 Z}{a}-\cphi\right) + 2 b \sin\left(\frac{2 Z}{a}+\cphi \right)\right), 
	\\ \nonumber 
	s_1^\cphi\left(r, \cphi, Z\right) & = & -2 \frac{S^Z_{101}}{3a} r^2 \left(b \cos\left(2 \cphi\right) - \frac{3}{4} r^2\right) - \frac{S^Z_{102}}{a} \left(b \cos\left(2 \cphi\right) - r^2\right) + S^\cphi_{10} - \\ \nonumber & & - S^r_{110} \frac{\sin\left(\cphi\right)}{r} + S^r_{120} \frac{\cos\left(\cphi\right)}{r} + \frac{S^Z_{21}}{20 a} r^4 \left(35 r^4 + 10 a^2 r^2+ 36 b^2 \cos\left(2 \cphi\right)^2 - \right. \\ \nonumber & & - \left. 80 b r^2 \cos\left(2 \cphi\right) + 12 b^2\right) + \frac{S^Z_{22}}{6a} r^2 \left[10 r^4+\left(3 a^2 + 6 b \right) r^2 + 4 b^2 \cos\left(2 \cphi\right)^2 - \right. \\ \nonumber & & \left. - 18 b r^2 \cos\left(2 \cphi\right) - 8 b^2 \cos\left(2 \cphi\right) + 4 b^2\right] + \frac{S^Z_{23}}{2 a} \left(\left(a^2 + 2 b\right) r^2 -2 b^2 \cos\left(2 \cphi\right) \right) - \\ \nonumber & & - \frac{S^r_{211}}{4 r} \left( \left(a^2 + 4 r^2 + 2 b\right) \sin\left(\frac{2 Z}{a}-\cphi\right) - 2 b \sin\left(\frac{2 Z}{a}+\cphi \right) \right) - \\ \nonumber & & - \frac{S^r_{212}}{4 r} \left(\left(a^2 + 4 r^2 + 2 b\right) \cos\left(\frac{2 Z}{a} - \cphi\right) - 2 b \cos\left(\frac{2Z}{a}+\cphi\right)\right), 
	\\ 
	s_1^Z\left(r, \cphi, Z\right) & = & \frac{S^Z_{101}}{4} r^4 + \frac{S^Z_{102}}{2} r^2 + S^Z_{103} - \frac{S^Z_{21}}{8} r^6 \left(12 b \cos\left(2 \cphi\right) - 7 r^2 + 4 b\right) - \\ \nonumber & & - S^Z_{22} r^4 \left(b \cos\left(2 \cphi\right) - \frac{5}{6} r^2\right) - \frac{b S^Z_{23}}{2} r^2 \cos\left(2 \cphi\right) - \\ \nonumber & & - \frac{a S^r_{211}}{2} r \sin\left(\frac{2 Z}{a} - \cphi\right) - \frac{a S^r_{212}}{2} r \cos\left(\frac{2 Z}{a} - \cphi\right), 
	\\ \nonumber 
	s_2^r\left(r, \cphi, Z\right) & = & S^r_{211} \cos\left(\frac{2 Z}{a} - \cphi\right) - S^r_{212} \sin\left(\frac{2 Z}{a} - \cphi\right),
	\\ \nonumber 
	s_2^\cphi\left(r, \cphi, Z\right) & = & \frac{1}{r} \left( S^r_{211} \sin\left(\frac{2 Z}{a} - \cphi\right) + S^r_{212} \cos\left(\frac{2 Z}{a} - \cphi\right) \right),
	\\ \nonumber 
	s_2^Z\left(r, \cphi, Z\right) & = & S^Z_{21} r^6 + S^Z_{22} r^4 + S^Z_{23} r^2 + S^Z_{24}.
\end{eqnarray}
However, substituting the solution~\eqref{an0cd0leadingsoln} in the remaining, lower order determining equations and proceeding with their solution, the compatibility of the first order equations for the derivatives of the function $m_2$, namely
$\partial_\cphi \left(\partial_r m_2(r, \cphi, Z) \right)= \partial_r \left(\partial_\cphi m_2(r, \cphi, Z)\right)$, immediately implies splitting into two cases:
\begin{itemize}
	\item $S^r_{211} = S^r_{212} = 0$, or
	\item $S^Z_{101} = S^Z_{23}$ and $S^Z_{21} = S^Z_{22} = 0$.	
\end{itemize}
In the first case, solving the remaining conditions, we find four systems, all already known:
\begin{itemize}
	\item three systems for which other choice of integrals of the standard cylindrical type is possible: the constant magnetic field and vanishing electrostatic potential, the constant magnetic field and isotropic harmonic oscillator in the plane perpendicular to the magnetic field~\cite{Marchesiello2022} and the superintegrable system~\eqref{oblateprolate_full_solution}; 
	\item the system with nonreducible generalized integrals~\eqref{cylell} defined in~\eqref{BellipAnotZero} and~\eqref{WellipAnotZero}, restricted to the particular values $c=d=0$.
\end{itemize}

In the second case, there is only one solution, namely the helical undulator of~\cite{KMS2022}, for which again other choices of commuting integrals exist, e.g. of standard Cartesian type.	

\subsection{$a=0$}\label{sec:azero-comp}
If $a=0$, equation \eqref{eq:split-a} is identically satisfied and \eqref{eqs2-1},\eqref{eqs2-2} and \eqref{eqs2-4} imply
\begin{equation}\label{Szsolcd}
	S_{21}^2(z)= -\frac{1}{2} d \partial_z S_{22}(z), \;\;S_{21}^{1}(z)= \frac{1}{2} c \partial_z S_{22}(z)
\end{equation}
and
\begin{eqnarray}
	&&c d \partial^2_{z}S_{22}(z)=0, \nonumber \\
	&& 4 b S_{22}(z)- c^2 \partial^2_{z}S_{22}(z)+ d^2 \partial^2_{z}S_{22}(z)=0. \label{eqcd-split}
\end{eqnarray}
This leads to a second splitting depending on whether $cd=0$ or not.

\subsubsection{$c$ or $d$ equals zero}\label{sec:dzero-comp}
Without loss of generality (i.e. up to a rotation and redefinition of the integrals) we can consider only the case $c$ not zero and $d=0$, therefore from \eqref{Szsolcd} we have $S_{21}^2(z)=0$ and the real solutions of equation \eqref{eqcd-split} are given by
\begin{equation}\label{S22Solbp}
	S_{22}(z)=\alpha_1 \cosh \left(\frac{2 \gamma z}{c}\right)+ \alpha_2 \sinh \left(\frac{2 \gamma z}{c}\right), \;\; \alpha_i\in\mathbb R
\end{equation}
for $ \gamma^2=b$, and
\begin{equation}\label{S22Solbn}
	S_{22}(z)=\alpha_1 \cos \left(\frac{2 \gamma z}{c}\right)+ \alpha_2 \sin \left(\frac{2 \gamma z}{c}\right), \;\; \alpha_i\in\mathbb R
\end{equation}
for $ \gamma^2=- b$. Let us continue with this last case, i.e. $S_{22}$ as in \eqref{S22Solbn}, for \eqref{S22Solbp} the computation is analogous. 

The last two equations \eqref{eqs2-5} and \eqref{eqs2-6} for the involutivity condition are solved by
\begin{eqnarray}
	s_1^1(x,y,z)&=&S_1^1(x,y)\nonumber\\
	&&-y\left( \sin \left(\frac{2 \gamma z}{c}\right) \left( \alpha_1\gamma x+\alpha_2\left(x^2+y^2- \gamma^2\right)\right)-\cos \left(\frac{2 \gamma z}{c}\right) \left(\gamma \alpha_2 x-\alpha_1 \left( x^2+y^2-\gamma^2\right)\right)\right),\nonumber\\
s_2^1(x,y,z)&=&S_1^2(x,y)
	+\frac{1}{4 \gamma}\left(\sin \left(\frac{2 \gamma z}{c}\right) \left(4\gamma\alpha_2 x (x^2+y^2)+\alpha_1(\gamma^2 x^2-c^2)\right)\right.\nonumber\\
	&&\left.+\gamma \cos \left(\frac{2 \gamma z}{c}\right) \left(4\gamma\alpha_1 x(x^2+y^2)-\alpha_2(\gamma^2 x^2-c^2)\right)\right). \label{SolS1azero}
\end{eqnarray}

We substitute the solution we have so far for $s_j^2$, $s_3^1$ and $B_1$, $B_2$, i.e. \eqref{solX22S}, \eqref{S13sol} and \eqref{solX22B1}--\eqref{solX22B2} respectively, together with \eqref{S22Solbn} into the second order equations \eqref{eqs21-1}-\eqref{eqs21-6} for the integral $X_1$. These give equations for the yet unknown functions $S_2^3$, $B_3$, $S_j^1$, $j=1,2,3$ that can be solved without much difficulty in this case. We obtain
\begin{eqnarray}
	S_2^3(x,y) &= &\frac{S_{21}^3}{20 y^4}+\frac{S_{22}^3}{6 y^2}+S_{24}^3 y+
	S_{23}^3,\\
	S_1^1(x,y) & = & \frac{1}{120 c y^4}\left(-40 \eta_2 y^2 \left(\gamma^2+x^2\right)+4 S_{21}^3 \left(3 \left(\gamma^2+x^2\right)^2-4 \gamma^2 y^2\right)\right.\nonumber\\
	&& +30S_{24}^3 y^5 \left(3 \left(x^2+y^2\right)^2-4 \gamma^2 y^2\right)+c^2 \left(10 S_{22}^3 y^2+3 S_{21}^3\right)\nonumber\\
	&&\left. +120 c y^4 (\omega_2 y+\omega_0)+120 \eta_1 y^5 \left(x^2+y^2\right)\right),
\end{eqnarray}
\begin{eqnarray}
	S_2^1(x,y) & = &\omega_1-\omega_2 x+ \frac{1}{2} c S_{24}^3 x+ \frac{x}{60 c y^3}\left(20 y^2 \left(-3 \eta_1 y \left(2 \gamma^2+x^2+y^2\right)-2 \eta_2\right)\right.\nonumber\\
	&&\left.+8 S_{21}^3 \left(\gamma^2+x^2\right)-15 S_{24}^3 y^3 \left(4 \gamma^2 x^2+3 \left(x^2+y^2\right)^2\right)\right),
\end{eqnarray}
and
\begin{eqnarray}
B_3(x,y,z) & = &\frac{6 \gamma^2 S_{21}^3+30 \eta_1 y^5-10 \eta_2 y^2+45 S_{24}^3 x^2 y^5+6 S_{21}^3 x^2+45 S_{24}^3 y^7}{30 c y^5}\nonumber\\
	&&-\left(\alpha_1 \left(\cos \left(\frac{2 \gamma z}{c}\right)\right)+ \alpha_2 \sin \left(\frac{2 \gamma z}{c}\right)\right).
\end{eqnarray}
We continue with the solution of the first order equations. They read
\begin{eqnarray}
	&&-2 b^2 \partial_x W(x,y,z)-B_2(x,y,z) S_1^3(x,y,z)+B_3(x,y,z) S_1^2(x,y,z)\nonumber\\
	&&-c\partial_z W(x,y,z)+\partial_x m_1(x,y,z)-2 y^2 \partial_x W(x,y,z)+2 x y \partial_y W(x,y,z)=0,\\
	&&B_1(x,y,z) S_1^3(x,y,z)-B_3(x,y,z)S_1^1(x,y,z)\nonumber\\
	&&+\partial_y m_1(x,y,z)-2 x^2\partial_y W(x,y,z)+2 x y\partial_x W(x,y,z)=0,\\
	&&-B_1(x,y,z)S_1^2(x,y,z)+B_2(x,y,z) S_1^1(x,y,z)\nonumber\\
	&&-c \partial_x W(x,y,z)+\partial_z m_1(x,y,z)=0
\end{eqnarray}
and
\begin{eqnarray}
	&& B_2(x,y,z) S_2^3(x,y,z)-B_3(x,y,z) S_2^2(x,y,z)-\partial_x m_2(x,y,z)=0,\\
	&&B_3(x,y,z) S_2^1(x,y,z)-B_1(x,y,z) S_2^3(x,y,z)-\partial_y m_2(x,y,z)=0,\\
	&&B_1(x,y,z) S_2^2(x,y,z)-B_2(x,y,z) S_2^1(x,y,z)-\partial_z m_2(x,y,z)+2 \partial_z W(x,y,z)=0,
\end{eqnarray}
for $X_1$ and $X_2$, respectively. The magnetic field $\vec{B}$ and the functions $S_i^j$ are already constrained by the previous steps. 

We use compatibility conditions for the above equations, consequence of the fact that the second order mixed derivatives of each $m_i$ must be equal. In particular, by imposing that 
$\partial_ {x y} m_2=\partial_{y x} m_2$ we get (after substituting all we know about $B_j$ and $S_i^j$)
\begin{eqnarray}
	&& -\frac{1}{10 c y^6}\left(\cos \left(\frac{2 \gamma z}{c}\right)\left(10 \alpha_1 S_{21}^3 x^3-10 \gamma \alpha_2 S_{21}^3 x^2\right.\right.\nonumber\\
	&&\left.+2 \alpha_1 x \left(5 \gamma^2 S_{21}^3+y^2 (2 S_{21}^3-5 \eta_2)\right)+\gamma \alpha_2 \left(y^2 \left(10 \eta_2-20 S_{23}^3 y^4\right.\right.\right.\nonumber\\
	&&\left.\left.\left.+S_{21}^3\right)-10 \gamma^2 S_{21}^3\right)\right)\nonumber\\
	&&+\left(10 \gamma^3 \alpha_1 S_{21}^3+10 \gamma^2 \alpha_2 S_{21}^3 x+2 \alpha_2 x \left(y^2 (2 S_{21}^3-5 \eta_2)+5 S_{21}^3 x^2\right)\right.\nonumber\\
	&&+\left.\left.\gamma \alpha_1 \left(-y^2 (10 \eta_2+S_{21}^3)+10 S_{21}^3 x^2+20 S_{23}^3 y^6\right)\right)\sin \left(\frac{2 \gamma z}{c}\right)\right)=0.
\end{eqnarray}
The above equation is polynomial in $x$ and $y$. By collecting the coefficients of different powers of $x$ and $y$, we see that necessarily
\begin{equation}
	10 S_{21}^3 \left( \alpha_2 \sin \left(\frac{2 \gamma z}{c}\right)+ \alpha_1 \cos \left(\frac{2 \gamma z}{c}\right) \right)=0.
\end{equation}
This leads to two subcases depending on $S_{21}^3$ vanishing or not. We find that there can be a solution corresponding to a new integrable system only for $S_{21}^3=S_{22}^3=S_{23}^3=\eta_2=\omega_0=0$. Namely, we obtain the integrable system given in subsection \ref{sec:ellipAZero}. 
The constants are for simplicity renamed according to $\gamma= \frac{c \delta}{2}$, $\eta_1=\beta_1$, $S_{24}^3=\beta_2$.

\subsubsection{$cd\neq0$}\label{sec:cdnotzero-comp}
If $cd\neq 0$ then the equations~\eqref{eqcd-split} imply that $S_{22}(z)$ vanishes and thus also $S_{21}^2(z)$ and $S_{22}^1(z)$ due to equation~\eqref{Szsolcd}.

Consequently, the equations \eqref{solX22S}--\eqref{solX22B2} simplify to 
\begin{eqnarray}
	s^1_2(x, y, z) & = & 0, \qquad s^2_2(x, y, z) = 0, \qquad s^3_2(x, y, z) = S_2^3(x, y), \nonumber \\
	B_1(x, y, z) & = & -\frac{1}{2} \partial_y S_2^3(x, y), \qquad B_2(x, y, z) = \frac{1}{2} \partial_x S_2^3(x, y). 
\end{eqnarray}
The remaining equations in~\eqref{eqs2-1}--\eqref{eqs2-6} imply 
\begin{equation}
	s^1_1(x, y, z) = S^1_1(x,y), \qquad s^2_1(x, y, z) = S^2_1(x,y), \qquad s^3_1(x, y, z) = S_3^1(x, y).
\end{equation}
Proceeding to solve the equations~\eqref{eqs21-1}--\eqref{eqs21-6} we determine the third component of the magnetic field using~\eqref{eqs21-2},
\begin{equation}
	B_3(x, y, z) = \frac{1}{4 xy} \left( d \partial_y S_2^3(x, y) - 2 \partial_y S_1^2(x, y) \right).
\end{equation}
The equation~\eqref{eqs21-3} then implies
\begin{equation}\label{cDxSdDyS}
	c \partial_x S_2^3(x, y) + d\partial_y S_2^3(x, y)=0.
\end{equation}
We can solve the remaining equations in~\eqref{eqs21-1}--\eqref{eqs21-6} with respect to the derivatives 
$\partial_y S_1^2(x, y)$, $\partial S_1^2(x, y)$, $\partial_y S_1^3(x, y)$, $\partial_x S_1^3(x, y)$ and consider their compatibility, namely
\begin{equation}
	\partial_y \left( \partial_x S_1^2(x, y) \right)=\partial_x \left( \partial_y S_1^2(x, y) \right), \qquad \partial_y \left( \partial_x S_1^3(x, y) \right)=\partial_x \left( \partial_y S_1^3(x, y) \right).
\end{equation}
These together with equation~\eqref{cDxSdDyS} define a system of linear partial differential equations for $S_2^3(x, y)$ and $S_1^1(x, y)$. Considering its differential consequences and their compatibility we find that we must have
\begin{eqnarray}
	S_1^1(x, y) & = & S^1_{11} y + S^1_{12} + \frac{S^1_{13}}{2} y (x^2 + y^2) + \frac{S^3_{21}}{4 c} \left(3 y (x^2 + y^2)^2 + 4 b y^3 - 2 c d x \right), \nonumber \\ 
	S_2^3(x, y) & = & S^3_{21} \left( y - \frac{d}{c} x \right) + S^3_{20}.
\end{eqnarray}
Next, integrating the equations determining the first order derivatives $\partial_y S_1^2(x, y)$, $\partial S_1^2(x, y)$, $\partial_y S_1^3(x, y)$, $\partial_x S_1^3(x, y)$ we find the complete solution of the second order conditions~\eqref{eqs21-1}--\eqref{eqB2} and~\eqref{eqs2-1}--\eqref{eqs2-6} in the form

\begin{eqnarray}\label{a0cdn0Bs}
	\nonumber B_1(x, y, z) & = & - \frac{S^3_{21}}{2}, \quad B_2(x, y, z) = - \frac{d S^3_{21}}{2c} , \quad B_3(x, y, z) = \frac{3 S^3_{21}}{2c} ( x^2 + y^2 ) + \frac{S^1_{13}}{2}, \\
	\nonumber s_1^1(x, y, z) & = & \frac{S^3_{21}}{4c} \left( 3 y \left( x^2 + y^2 \right)^2
	+ 2 (2 b y^3 - c d x) \right) +\frac{S^1_{13}}{2} y \left( x^2 + y^2 \right) +S^1_{11} y +S^1_{12}, \\
	s_1^2(x, y, z) & = & - \frac{S^3_{21}}{4c} \left( 3 x \left(x^2+y^2 \right)^2 + 2 \left( \left( d^2 - c^2 \right) x - 2 b x^3 - c d y \right) \right) \\ & & 
	\nonumber - \frac{S^1_{13}}{2} x \left(x^2 + y^2 \right)+ \left( S^1_{13} b - S^1_{11} \right) x + S^2_{1},
	\\
	\nonumber s_1^3(x, y, z) & = & \frac{S^3_{21}}{2} \left( \left( y - \frac{d}{c} x \right) \left(x^2 + y^2\right) + 2 b \frac{d}{c} x \right) + \frac{S^1_{13}}{2} c \left( y - \frac{d}{c} x \right)+S_1^3, \\
	\nonumber s_2^1(x, y, z) & = & 0, \qquad s_2^2(x, y, z) = 0, \qquad s_2^3(x, y, z) = S^3_{21} \left( y - \frac{d}{c} x \right) + S^3_{20}.
\end{eqnarray}

Proceeding to solve the lower order conditions, we immediately find from the involutivity condition $\{ X_1,X_2 \}=0$ that $W(x,y,z)$ can depend only on the $xy$--coordinates and express $\partial_x W(x, y)$ in terms of $\partial_y W(x, y)$ and the integration constants $S^3_{21}, S_{13}^1,\ldots$ introduced in~\eqref{a0cdn0Bs}. Substituting it into the compatibility conditions $\partial_u \left( \partial_v m_a\right)= \partial_v \left( \partial_u m_a\right)$, $a=1,2$, $u,v=x,y,z$ derived from the first order conditions coming from $\{H,X_1\}=0$ and $\{H,X_2\}=0$, and easily integrating them, we find $W(x,y)$ fully determined up to an irrelevant additive constant. However, comparing it with the previously derived results, we see that this system is the limit of the system~\eqref{BellipAnotZero}--\eqref{WellipAnotZero} as $a\rightarrow 0$, $a \beta_1 \rightarrow \frac{S^1_{13}}{2}$, $\beta_2=-\frac{S^3_{21}}{2}$, $\omega_1=-\frac{S^2_1}{c}$, $\omega_2=\frac{S^1_{12}}{c}$, $\omega_3=\frac{S^1_{11}}{2 c}$. 

\begin{acknowledgement}
\noindent {\bf Funding:} FH was supported by the project grant CZ.02.2.69/0.0/0.0/18\_053/0016980 Mobility CTU - STA, Ministry of Education, Youth and Sports of the Czech Republic, co-financed by the European Union. OK was supported by the Grant Agency of the Czech Technical University in Prague, grant No. SGS22/178/OHK4/3T/14. AM acknowledges GNFM--INdAM for  support. L\v{S} was supported by the project of the Ministry of Education, Youth and Sports of the Czech Republic CZ.02.1.01/0.0 /0.0/16\_019/0000778 Centre of Advanced Applied Sciences, co-financed by the European Union.\smallskip

\noindent {\bf Declaration of Competing Interest Statement:} The authors declare that they have no known competing financial interests or personal relationships that could have appeared to influence the work reported in this paper.\smallskip

\noindent {\bf Data Availability Statement:} All relevant data generated or analyzed during this study are included in this article. 
\end{acknowledgement}

\pdfbookmark[1]{References}{ref}
\bibliographystyle{iopart-num}

\begin{thebibliography}{10}
\expandafter\ifx\csname url\endcsname\relax
  \def\url#1{{\tt #1}}\fi
\expandafter\ifx\csname urlprefix\endcsname\relax\def\urlprefix{URL }\fi
\providecommand{\eprint}[2][]{\url{#2}}

\bibitem{Makarov1967}
Makarov A, Smorodinsky Y, Valiev K and Winternitz P 1967 {\em Nuovo Cimento A
  Series\/} {\bf 10} 1061--1084
  \urlprefix\url{https://doi.org/10.1007/BF02755212}

\bibitem{FrisMandrosov}
Fri\v{s} J~Mandrosov V, Smorodinsky Y, Uhl\'{\i}\v{r} M and Winternitz P 1965
  {\em Phys. Lett.\/} {\bf 16} 354--356
  \urlprefix\url{https://doi.org/10.1016/0031-9163(65)90885-1}

\bibitem{WinternitzSmorodinski}
Winternitz P, Smorodinsky Y~A, Uhl\'{\i}\v{r} M and Fri\v{s} I 1967 {\em Soviet
  J. Nuclear Phys.\/} {\bf 4} 444--450

\bibitem{Evans}
Evans N~W 1990 {\em Phys. Rev. A\/} {\bf 41} 5666--5676
  \urlprefix\url{https://doi.org/10.1103/PhysRevA.41.5666}

\bibitem{BaShaMe}
Bagrov V~G, Shapovalov V~N and Meshkov A~G 1974 {\em Soviet Physics J.\/} {\bf
  15} 1115–1119 \urlprefix\url{https://doi.org/10.1007/bf00910289}

\bibitem{Benenti_2001}
Benenti S, Chanu C and Rastelli G 2001 {\em J. Phys. A\/} {\bf 42} 2065--2091
  \urlprefix\url{https://doi.org/10.1063/1.1340868}

\bibitem{Marchesiello2022}
Marchesiello A and {\v{S}}nobl L 2022 {\em J. Phys. A\/} {\bf 55} 145203
  \urlprefix\url{https://doi.org/10.1088/1751-8121/ac515e}

\bibitem{DoGraRaWin}
Dorizzi B, Grammaticos B, Ramani A and Winternitz P 1985 {\em J. Math. Phys.\/}
  {\bf 26} 3070--3079 \urlprefix\url{http://dx.doi.org/10.1063/1.526685}

\bibitem{McSWin}
McSween E and Winternitz P 2000 {\em J. Math. Phys.\/} {\bf 41} 2957--2967
  \urlprefix\url{http://dx.doi.org/10.1063/1.533283}

\bibitem{BeWin}
B{\'e}rub{\'e} J and Winternitz P 2004 {\em J. Math. Phys.\/} {\bf 45}
  1959--1973 \urlprefix\url{http://dx.doi.org/10.1063/1.1695447}

\bibitem{Pucacco_2005}
Pucacco G and Rosquist K 2005 {\em J. Math. Phys.\/} {\bf 46} 012701
  \urlprefix\url{{https://doi.org/10.1063/1.1818721}}

\bibitem{KMS2022}
Kub\r{u} O, Marchesiello A and \v{S}nobl L 2023 {\em Ann. Phys.\/} {\bf 451}
  169264 \urlprefix\url{https://doi.org/10.1016/j.aop.2023.169264}

\bibitem{HS2022}
Hoque F and \v{S}nobl L 2023 {\em J. Phys. A\/} {\bf 56} 165203
  \urlprefix\url{https://doi.org/10.1088/1751-8121/acc55a}

\bibitem{Marchesiello2015a}
Marchesiello A, Post S and {\v{S}}nobl L 2015 {\em J. Math. Phys.\/} {\bf 56}
  102104 \urlprefix\url{https://doi.org/10.1063/1.4933218}

\bibitem{Miller2013}
Miller Jr W, Post S and Winternitz P 2013 {\em J. Phys. A\/} {\bf 46} 423001,
  97 \urlprefix\url{https://doi.org/10.1088/1751-8113/46/42/423001}

\bibitem{Marchesiello2017}
Marchesiello A and \v{S}nobl L 2017 {\em J. Phys. A\/} {\bf 50} 245202, 24
  \urlprefix\url{https://doi.org/10.1088/1751-8121/aa6f68}

\bibitem{BKS2020}
Bertrand S, Kub{\r{u}} O and {\v{S}}nobl L 2020 {\em J. Phys. A: Math.
  Theor.\/} {\bf 54} 015201
  \urlprefix\url{https://doi.org/10.1088/1751-8121/abc4b8}

\bibitem{Zhang}
Zhang P~M, Zou L~P, Horvathy P and Gibbons G 2014 {\em Ann. Phys.\/} {\bf 341}
  94--116 \urlprefix\url{https://doi.org/10.1016/j.aop.2013.11.004}

\bibitem{Marchesiello2015}
Marchesiello A, \v{S}nobl L and Winternitz P 2015 {\em J. Phys. A\/} {\bf 48}
  395206, 24 \urlprefix\url{https://doi.org/10.1088/1751-8113/48/39/395206}

\bibitem{BertrandSnobl}
Bertrand S and \v{S}nobl L 2019 {\em J. Phys. A\/} {\bf 52} 195201, 25
  \urlprefix\url{https://doi.org/10.1088/1751-8121/ab14c2}

\bibitem{Fournier2019}
Fournier F, \v{S}nobl L and Winternitz P 2020 {\em J. Phys. A\/} {\bf 53}
  085203 \urlprefix\url{https://doi.org/10.1088/1751-8121/ab64a6}

\bibitem{Marchesiello2018Sph}
Marchesiello A, \v{S}nobl L and Winternitz P 2018 {\em J. Phys. A\/} {\bf 51}
  135205, 24 \urlprefix\url{https://doi.org/10.1088/1751-8121/aaae9b}

\bibitem{BALAL2020163895}
Balal N, Bratman V and Magory E 2020 {\em Nucl. Instrum. Methods Phys. Res.\/}
  {\bf 971} 163895 \urlprefix\url{https://doi.org/10.1016/j.nima.2020.163895}

\bibitem{HeinIld}
Heinzl T and Ilderton A 2017 {\em J. Phys. A\/} {\bf 50} 345204, 14
  \urlprefix\url{https://doi.org/10.1088/1751-8121/aa7fa3}

\bibitem{LACROIX2009906}
Lacroix C, Taillefumier M, Dugaev V, Canals B and Bruno P 2009 {\em J. Magn.
  Magn. Mater.\/} {\bf 321} 906--908
  \urlprefix\url{https://doi.org/10.1016/j.jmmm.2008.11.085}

\bibitem{SINITSYN200225}
Sinitsyn N 2002 {\em J. Magn. Magn. Mater.\/} {\bf 253} 25--27
  \urlprefix\url{https://doi.org/10.1016/S0304-8853(01)00694-1}

\bibitem{HUSE_Optik}
Huse V, Sharma G, Mishra S and Mishra G 2014 {\em Optik\/} {\bf 125} 4739--4741
  \urlprefix\url{https://doi.org/10.1016/j.ijleo.2014.04.071}

\bibitem{Kubu_2021}
Kub{\r{u}} O, Marchesiello A and {\v{S}}nobl L 2021 {\em J. Phys. A: Math.
  Theor.\/} {\bf 54} 425204
  \urlprefix\url{https://doi.org/10.1088/1751-8121/ac2476}

\end{thebibliography}
\providecommand{\newblock}{}

\end{document}